\documentclass{aa}
\pdfoutput=1
\usepackage{graphicx}
\usepackage{amsmath}	
\usepackage{amssymb}	
\usepackage{multicol}        
\usepackage{bm}		
\usepackage{pdflscape}	
\usepackage{xcolor}
\usepackage{esint}

\newcommand{\vsini}{$v \sin i$}
\newcommand{\chit}{$\chi^{2}$}
\newcommand{\halpha}{H$\alpha$}
\newcommand{\teff}{$T_\mathrm{eff}$}
\newcommand{\FeH}{[Fe/H]}

\newcommand{\avB}{$\langle B_V \rangle$~}
\newcommand{\RXJ}{LQ Lup~}
\newcommand{\logg}{$\log g~$}
\newcommand{\StI}{Stokes \textit{I}}
\newcommand{\StV}{Stokes \textit{V}}
\def\kms{\hbox{km\,s$^{-1}$}}  
  
\def\rpd{\hbox{rad\,d$^{-1}$}}

\begin{document}
\title{Magnetic field and prominences of the young, solar-like, ultra-rapid rotator V530 Per}
\author{T.-Q. Cang \inst{1} \and P. Petit \inst{1} \and J.-F. Donati \inst{1} \and C.P. Folsom \inst{1} \and M. Jardine \inst{2} \and C. Villarreal D’Angelo \inst{3} \and \\A.A. Vidotto \inst{3} \and S.C. Marsden \inst{4} \and F. Gallet \inst{5} \and B. Zaire \inst{1}}

\institute{Institut de Recherche en Astrophysique et Plan\'etologie, Universit\'e de Toulouse, CNRS, CNES, 14 avenue Edouard Belin, 31400 Toulouse, France \\
\email{Tianqi.Cang@irap.omp.eu}
\and
SUPA, School of Physics and Astronomy, University of St Andrews, North Haugh, KY16 9SS, UK
\and
School of Physics, Trinity College Dublin, the University of Dublin, Dublin-2, Ireland
\and
University of Southern Queensland, Centre for Astrophysics, Toowoomba, 4350, Australia
\and
Univ. Grenoble Alpes, CNRS, IPAG, 38000 Grenoble, France}

\abstract
{Young solar analogs reaching the main sequence experience very strong magnetic activity, generating angular momentum losses through wind and mass ejections.}
{We investigate signatures of magnetic fields and activity at the surface and in the prominence system of the ultra-rapid rotator V530 Per, a G-type solar-like member of the young open cluster $\alpha$~Persei. This object has a rotation period that is shorter than all stars with available magnetic maps.} 
{With a time-series of spectropolarimetric observations gathered with ESPaDOnS over two nights on the Canada-France-Hawaii Telescope (CFHT), we reconstructed the surface brightness and large-scale magnetic field of V530 Per using the Zeeman-Doppler imaging method, assuming an oblate stellar surface. We also estimated the short term evolution of the brightness distribution through latitudinal differential rotation. Using the same data set, we finally mapped the spatial distribution of prominences through tomography of the \halpha\ emission.}
{The brightness map is dominated by a large, dark spot near the pole, accompanied by a complex distribution of bright and dark features at lower latitudes. Taking the brightness map into account, the magnetic field map is reconstructed as well. Most of the large-scale magnetic field energy is stored in the toroidal field component. The main radial field structure is a positive region of about 500 G, at the location of the dark polar spot. The brightness map of V530 Per is sheared by solar-like differential rotation, with roughly a solar value for the difference in rotation rate between the pole and equator. It is important to note that \halpha~is observed in emission and it is mostly modulated by the stellar rotation period over one night. The prominence system is organized in a ring at the approximate location of the corotation radius, and displays significant evolution between the two observing nights.}
{V530 Per is the first example of a solar-type star to have its surface magnetic field and prominences mapped together, which will bring important observational constraints to better understand the role of slingshot prominences in the angular momentum evolution of the most active stars.}

\maketitle

\section{Introduction}

A large fraction of young Suns close to the early main sequence experience very large rotation rates, as they still possess most of the angular momentum acquired during the stellar formation process (see the review of \citealt{2013EAS....62..143B}). This type of rapid rotation is responsible for the efficient amplification of internal magnetic fields through the action of a global dynamo, as observed in most indirect activity tracers. This is the case, for example, in Ca II H\&K emission \citep{1984ApJ...279..763N}, X-ray flux \citep{2011ApJ...743...48W}, or photometric variability due to spots \citep{2013MNRAS.436.1883W} and flares \citep{2016ApJ...829...23D}. However, for stars with sufficiently large rotation rates, magnetic activity seems to reach an upper limit although the exact rotation threshold for saturation depends on the magnetic tracer taken into account. X-ray observations clearly highlight this so-called saturation phenomenon  \citep{1981ApJ...248..279P, 1996AJ....112.1570P,2011ApJ...743...48W}, as well as Zeeman broadening measurements \citep{2012LRSP....9....1R} or large-scale magnetic field measurements \citep{2014MNRAS.441.2361V,2019ApJ...876..118S}. Attempts to model this saturated state in global numerical simulations of G-K stars remain scarce (e.g.,  \citealt{2017EPJWC.16002010A, 2019ApJ...880....6G}).

Tomographic mapping is a powerful approach to characterize the large-scale surface magnetic fields of rapid rotators. Since its first application to an active solar-type star with HR 1099 \citep{1992A&A...265..682D}, Zeeman-Doppler imaging (ZDI hereafter) has been applied to several dozens of cool active stars on the main sequence (e.g.,  \citealt{2008MNRAS.388...80P} for solar analogs, \citealt{2008MNRAS.390..567M} for M dwarfs, or \citealt{2019ApJ...876..118S} for a global study). Several recent ZDI studies have specifically investigated how magnetic geometries of Sun-like stars evolve during the early main sequence \citep{2014MNRAS.441.2361V, 2016MNRAS.457..580F,2016A&A...593A..35R, 2018MNRAS.474.4956F}. Although the first cool ZDI targets were saturated stars as their Zeeman signatures are easier to detect, the most recent observing projects dealing with G-K stars have concentrated on objects in the unsaturated regime. As a consequence, while the unsaturated dynamo regime is now well sampled by the ZDI models available so far, we are still left with few G-K dwarfs in the saturated regime, which is mostly populated by M dwarfs in ZDI surveys. Our observations of v530~Per are aimed at enlarging the sample of fast rotators studied with ZDI.

Fast rotators are also ideal laboratories for studying stellar prominences and their impact on angular momentum evolution in young stars. Prominences are dense clouds of gas at chromospheric temperature, trapped in closed stellar magnetic loops and extending into the hot and tenuous corona. Prominences, along with stellar winds, remove angular momentum from stars and therefore contribute to the early evolution of active stars \citep{2012ApJ...760....9A,2019MNRAS.tmp.2779J}. For rapidly rotating stars, prominence systems become much more massive and extended than on the Sun
\citep[see the review by][]{1999ASPC..158..146C}.  Observational signatures of stellar prominences are usually extracted from Balmer lines. They show up as absorption features in the line profile when the prominence transits in front of the stellar disk, and generate line emission otherwise. Following a first detection by \cite{1989MNRAS.236...57C}, prominence systems have been reported in a small number of active G and K stars \citep{2000MNRAS.316..699D,2006MNRAS.365..530D}, M-dwarfs \citep{1996A&A...311..651B,1998A&A...337..757E}, and PMS stars \citep{2008MNRAS.385..708S,2009MNRAS.399.1829S}. When spectral signatures are seen in emission (i.e., mainly if the line-of-sight inclination of the stellar rotation axis is small), the spatial distribution of prominences can be reconstructed through tomographic models inspired from the observation of cataclysmic variables \citep{2000MNRAS.316..699D,2001MNRAS.326.1057B}.

Here we investigated the large-scale photospheric magnetic field and prominence system of V530~Per (also named AP 149), which is a cool, rapidly rotating member of the young open cluster $\alpha$ Persei \citep{1992AJ....103..488P}. Combining its X-ray flux and projected rotational velocity, V530 Per was proposed to be a saturated \citep{1994MNRAS.268..181O} or even super-saturated \citep{1996AJ....112.1570P} star, making it an interesting object for ZDI studies lacking G-K stars in this extreme magnetic regime. Its sustained magnetic activity is also responsible for regular photometric variations attributed to a $\sim$8~hr rotation period \citep{1993MNRAS.262..521O}. Doppler mapping performed by \citet{2001MNRAS.326.1057B} revealed the presence of a large, dark spot near the visible rotation pole. They also recovered a first prominence map from their \halpha~time series, unveiling large prominences extending up to several stellar radii.

In this paper, we presented a time-series of spectropolarimetric observations of V530 Per (Sec. \ref{sec:obs}). We first refine and discuss its fundamental parameters (Sec. \ref{sec:param}) and then reconstruct its brightness and magnetic field map (Sec. \ref{sec:ZDI}). We also modeled the latitudinal differential rotation of V530 Per (Sec. \ref{sec: df}) and present two prominence maps reconstructed from two distinct nights (Sec. \ref{sec:prom}). Finally, we discuss our results in the light of previous works (Sec. \ref{sec:disc}).

\section{Observational material}
\label{sec:obs}


\begin{table}
\caption{Observation log of V530 Per for 29 Nov \& 05 Dec 2006. From left to right, we list the date, the Julian date, the rotational phases calculated with Eq. \ref{equ:emp}, and the peak S/N.}
          \centering
          \begin{tabular}{@{} cccc @{}}
            \hline
Date (2006)  & HJD (2454000+) & $E$ & peak S/N\\ 
\hline
29Nov & 69.71271 &-7.1366 &94  \\
29Nov & 69.74455 &-7.0373 &103 \\
29Nov & 69.77677 &-6.9368 &103 \\
29Nov & 69.80806 &-6.8391 &113 \\
29Nov & 69.84102 &-6.7363 &116 \\
29Nov & 69.87316 &-6.6360 &109 \\
29Nov & 69.91170 &-6.5158 &130 \\
\hline
05Dec & 75.70177 &11.5500 &105 \\
05Dec & 75.73473 &11.6528 &117 \\
05Dec & 75.76595 &11.7502 &119 \\
05Dec & 75.79718 &11.8477 &111 \\
05Dec & 75.83026 &11.9509 &107 \\
05Dec & 75.86149 &12.0483 &106 \\
05Dec & 75.89713 &12.1595 &113 \\
            \hline
          \end{tabular}  
          \label{tab:obslog}
\end{table}

We obtained a time-series of spectropolarimetric observations of V530 Per in late 2006, over two nights separated by a 6-day gap (November 29 and December 05). The data were collected at Mauna Kea observatory by the ESPaDOnS spectropolarimeter \citep{2006ASPC..358..362D}, mounted at the Cassegrain focus of the Canada-France-Hawaii Telescope (CFHT). We used the polarimetric mode of this instrument, delivering a spectral resolution of about 65,000 and simultaneous coverage of the wavelength domain between 0.37 and 1.05 $\mathrm{\mu m}$. Circular polarization sequences (\StV parameter) were collected as part of our program, as Zeeman signatures are much larger in this polarization state (a factor of $\sim$10 stronger than $Q~\&~U$, e.g., \citealt{1992soti.book...71L,2000NewA....5..455W,2011ApJ...732L..19K}). Every polarization sequence consists of four subexposures with a fixed integration time of 600s, and different angles of the two half-wave rotatable Fresnel rhombs in the polarimetric module, following a procedure designed to remove spurious polarization signatures at first order \citep{1993A&A...278..231S}. Normalized, reduced \StI~and \textit{V} spectra are extracted from the raw ESPaDOnS images using the \texttt{Libre-ESpRIT} automatic pipeline tool \citep{1997MNRAS.291..658D,2006ASPC..358..362D}. The typical peak signal-to-noise ratio (S/N) of our \StV~spectra is slightly above 100, while \StI~spectra corresponding to single subexposures have a peak S/N of about 50. In the rest of this study, all tasks involving \StI~spectra alone make use of the subexposures, as they offer a denser temporal sampling. All reduced spectra analyzed here are available through the PolarBase archive \citep{2014PASP..126..469P}.

We obtained an equal number of observations during both telescope nights, leading to a total of 14 \StV~spectra and 56 \StI~spectra. To assign a rotational phase ($E$) to every observation, we used the ephemeris:
\begin{equation}
\label{equ:emp}
\mathit{HJD}_{\mathrm{obs}} = \mathit{HJD}_{0} + P_{\mathrm{rot}} \times E
\end{equation}
\noindent where $P_{\mathrm{rot}}=0.3205$~d is the rotational period of the equator, taken from our differential rotation measurement (see Sec. \ref{sec:DR}), and the initial Heliocentric Julian date $\mathit{HJD}_{0} = 2454072.0$ is arbitrarily selected between the two observing nights. The resulting phases, reported in Table \ref{tab:obslog}, show that successive \StV~spectra are separated by about 10\% of a rotation cycle. The phase smearing during the collection of a \StV~sequence, which is of the same order, may be responsible for a reduced amplitude of polarized signatures generated by low latitude features (experiencing the largest Doppler shifts). Our observations in each individual night were able to cover about $60\%$ of one rotation cycle. The rotational phase reached at the end of the first night was within 10\% of the phase observed at the beginning of the second night. We therefore end up with a complete phase coverage of the target, and with redundant observations between phases 0.85 and 0.15. 

\section{Fundamental parameters of V530 Per}
\label{sec:param}

\begin{figure} 
\centering
    \includegraphics[width=1.0\linewidth]{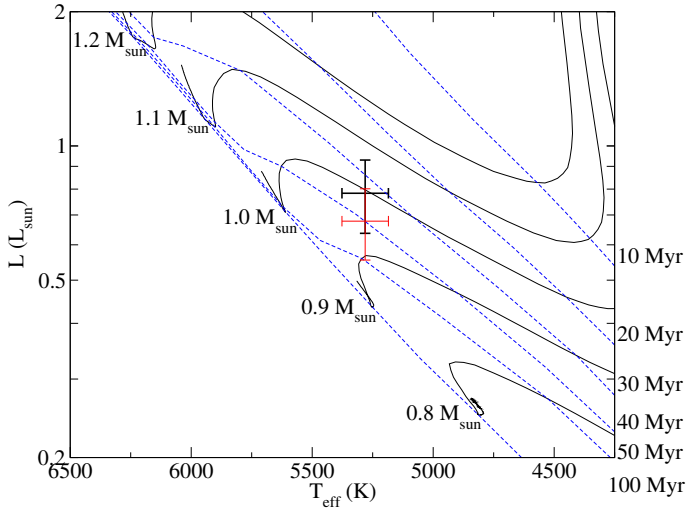}
    \caption{V530 Per in the Hertzsprung-Russell diagram, with evolutionary tracks generated by the {\sc starevol} code. Evolutionary tracks are given for 0.1 $M_\odot$ steps (full lines), and isochrones are overimposed for 10, 20, 30, 40, 50 and 100 Myr (dashed lines). The black and red crosses are obtained from the K and J magnitudes, respectively.}
\label{fig:tefflogg}
\end{figure}


\begin{table}[ht]

\caption{Fundamental parameters of V530 Per}
          \centering
          \begin{tabular}{lll}
        \hline
 Name & Value&  References\\ 
\hline
Distance & $167.7\pm15$ pc& 1,2\\
Age  &$33^{+10}_{-7}$~Myr& 1 \\
$T_\mathrm{eff}$ &$5281\pm96$ K& 1\\
$\log g$ &$4.4 \pm 0.1$  [cm~s$^{-2}$]& 1\\
$[Fe/H]$ & $-0.16 \pm 0.08$ & 1\\
$m_V^{min}$ &$11.657\pm0.13$&3\\
$m_J$ & $10.08 \pm 0.019 $ & 4 \\
$m_K$  &$9.422 \pm0.019 $& 4 \\
Luminosity & $0.78 \pm 0.15~L_{\sun}$ &1\\
Radius & $1.06 \pm 0.11~R_{\sun}$ & 1 \\
Mass & $1.00\pm0.05$~$M_{\sun}$ &  1 \\
$\log L_{x}$ & 31.2  [erg~s$^{-1}$]  &5 \\ %
Rossby number &$0.013 \pm0.002$& 1\\
Co-rotation radius &$1.9 \pm 0.2 ~R_{*}$& 1\\
Alfv\'en radius & $5~R_{*}$ &1\\
\vsini & $106 \pm 1~\kms$ & 1  \\
Eq. rot. period & $0.32055 \pm 0.00005$~d& 1\\
$d\Omega$ & $0.042 \pm 0.005$~\rpd& 1\\
Inclination angle & $40\pm4^{\circ}$  & 1 \\
Radial velocity  & $-0.96\pm0.1~\kms $ & 1 \\
        \hline
        \end{tabular}
\textbf{\tt{References: }}~1. This work 2. \cite{2018A&A...615A..12Y} 
3. \cite{2013AJ....145...44Z} 4.\cite{2003tmc..book.....C} 5. \cite{2013AJ....145..143P}\\

\label{tab:para}
\end{table}

V530 Per is a member of the $\alpha$~Persei open cluster \citep{1992AJ....103..488P}, a relatively young open cluster with an age of $63^{+8}_{-27}$ Myr, as derived from GAIA DR1 data \citep{2018A&A...615A..12Y}. This recent estimate is significantly smaller than the $90\pm10$ Myr reported by \cite{1999ApJ...527..219S}. The global metallicity of the cluster is close to solar, with [Fe/H]=$-0.10 \pm 0.08$ reported by \citet{2011MNRAS.410.2526B}, who further noticed that stars with \teff > 5500~K have [Fe/H] $\approx -0.04$, while stars with \teff < 5500 K feature a lower metallicity with [Fe/H] $\approx-0.13$. 

Since there is no available distance measurement for V530 Per itself, we used the average distance of the cluster in our work. Using the GAIA DR1 catalog, \citet{2018A&A...615A..12Y} derived an average distance $d=167.7\pm0.3$~pc. This value is smaller than the one derived from Hipparcos parallax ($172.4\pm2.7$ pc, \citealt{2009A&A...497..209V}). Considering individual stellar parallaxes reported by \cite{2018A&A...615A..12Y}, we derived a standard deviation of 0.46 mas for $\alpha$~Per members, that we considered to be the cluster extent. We adopted this value as our uncertainty on the parallax of V530 Per, translating into a conservative distance uncertainty of about 15 pc.

We used our high-resolution spectral data to measure surface fundamental parameters of V530 Per (such as its surface temperature, gravity or metallicity) which are not documented in the literature. We mostly repeated here the procedure detailed by \cite{2016MNRAS.457..580F} and already applied to a sample of young solar-type stars \citep{2016MNRAS.457..580F,2018MNRAS.474.4956F}. 
This approach iteratively fits synthetic spectra to the observation by $\chi^2$ minimization.  We computed spectra with the {\sc zeeman} spectrum synthesis code \citep{1988ApJ...326..967L,Wade2001-zeeman}, which performs polarized radiative transfer in LTE and works well for stars as cool as 5200 K (e.g. \citealt{2018MNRAS.474.4956F}).  We used {\sc marcs} model atmospheres \citep{Gustafsson2008} as input, together with atomic data extracted from the VALD database \citep{1997BaltA...6..244R,1999A&AS..138..119K,Ryabchikova2015-VALD3}.  This approach using {\sc Zeeman} has been verified against alternate spectroscopic parameter determinations \citep{2016MNRAS.457..580F,2018MNRAS.474.4956F}, and an interferometric determination \citep{2018MNRAS.481.5286F} for stars in this range of spectral types, with good agreement consistently found. One should note that we used single spectra for the parameter determination. The S/N is sufficiently high that it is not the limiting factor on our results, and given the heavily spotted nature of the star, any parameter determination assuming a uniform atmosphere will necessarily be approximate.

The observed spectrum of V530 Per was first renormalized, with the synthetic spectra providing guidelines for regions best approximating the continuum. The  theoretical spectra were then compared to the renormalized spectrum, focusing on several spectral regions that are mostly uncontaminated by strong molecular lines, which are not taken into account by {\sc Zeeman}, and telluric lines. 

A first fit was performed assuming a solar metallicity, which is a reasonable approximation since 
$\alpha$ Per members are known to have near solar metallicity \citep{2011MNRAS.410.2526B}. This fit was performed using 5 spectral regions $\sim$10 nm long, between 600 and 650~nm (specifically 600-610, 607.5-620.5, 619.6-627.55, 631.2-634.1 + 635.0-640.4, and 640.4-644.6 + 645.9-650.4~nm). 
The average of the best fits for individual windows was taken as the final value, and the standard deviation was taken as an uncertainty estimate.  

Assuming a solar metallicity, we obtained an effective temperature \teff $= 5281\pm96$~K, a surface gravity $\log g = 4.10\pm0.19$, a projected rotational velocity \vsini\ $ = 116.70 \pm 2.38$~\kms\ (consistent with the estimate of \citealt{2010MNRAS.402.1380J}, but significantly larger than the value of \citealt{2001MNRAS.326.1057B}) and a micro turbulence $v_{\rm mic} = 1.3 \pm 0.4$~\kms. 
Although we checked on less noisy data sets that spectra  of extremely active stars collected at different rotational phases do not produce significantly different results, except for \vsini\ estimates, our approach is still limited by the fact that we assumed the atmosphere to be homogeneous over the whole stellar surface. This is far from the actual situation of V530 Per, which is covered by a complex mixture of cool and hot spots (see Sec. \ref{sec:maps}). In particular, the giant, dark polar spot of this extremely active star impacts the line shape by generating a broad bump in the line bottom \citep{2001MNRAS.326.1057B}, which has the effect of biasing our \vsini\ estimate toward larger values. An independent estimate of \vsini, using Doppler mapping and incorporating the effect of surface spots in the model, leads to a significantly smaller value of about 105.6~\kms\  (Sec. \ref{sec: ZDIpara}). 
A second fit was performed with the metallicity left as a free parameter (as well as \teff, \logg, and \FeH), To better constrain the additional parameter, this round of fitting included an additional 5 spectral regions between 550 and 600 nm (specifically 550-560.7, 560.7-569.2, 569.2-580, 580-590.3, and 590.3-600 nm), to better constrain this parameter. These additional windows have more severe line blending and consequently an accurate continuum normalization is more difficult, but they improve the statistical validity and provide more data to better constrain parameters with similar effects on the spectrum. 
The outcome is a set of atmospheric parameters in good agreement with our previous estimate, and the metallicity is found to be [Fe/H] $= -0.16 \pm 0.08$. This is consistent with the average value of \cite{2011MNRAS.410.2526B}, although our estimate is sensitive to small departures from a perfect continuum normalization. 

The V magnitude of V530 Per varies with time, with reported values  between $11.657\pm0.13$ \citep{2013AJ....145...44Z} and $11.981\pm0.073$ \citep{2015AAS...22533616H}. Assuming that the brightness variations of V530 Per all come from rotating star spots, we considered the brightest available magnitude as the nonspotted magnitude. This magnitude was then used to estimate the luminosity $L=0.78\pm 0.18~L_{\sun}$, using the distance discussed above ($167.7 \pm 15$ pc), 
the V band bolometric correction 
$BC_{\rm v} = -0.23$ from \citet{2013ApJS..208....9P}
and the reddening correction $A_V = 0.312$ from \citet{1998ApJ...504..170P}. Subsequently, we derive the stellar radius $R = 1.06\pm0.11~R_{\sun}$.
We then repeated the same procedure with the K magnitude, as it is much less affected by activity-induced fluctuations \citep[e.g., ][]{2010RAA....10..253F} and suffers less from the interstellar extinction. Using a K magnitude of $9.422 \pm 0.019$ from \citet{2003tmc..book.....C}, and the interpolated bolometric correction $BC_{K} = 1.706 \pm 0.056$ of \citet{2006A&A...450..735M}, we get $L=0.78\pm 0.15~L_{\sun}$ and $R = 1.06\pm0.11~R_{\sun}$. Alternately, using the J band magnitude and the bolometric correction $BC_J=1.22\pm 0.03$ from \citet{ 2013ApJS..208....9P}, we get $L=0.68 \pm 0.12~L_{\sun}$ and $R=0.98 \pm 0.10~R_{\sun}$, in good agreement with the K band estimate.  

Using our values of the effective temperature and luminosity, we obtained the HR diagram  of Fig. \ref{fig:tefflogg}. According to evolutionary tracks computed with the {\sc starevol} code by \citet{Amard2016-starevol-rot,Amard2019-starevol-tracks} for solar metallicity, we get $M = 1.00 \pm 0.05 ~M_{\sun}$, $\log g = 4.4 \pm 0.1$ and an age of $33^{+10}_{-7}$~Myr using the K band, versus $M = 0.95 \pm 0.05 ~M_{\odot}$, $\log g = 4.4 \pm 0.1$ and an age of $40^{+11}_{-8} $~Myr with the J band. Both ages are consistent, within uncertainties, with the $63^{+8}_{-27}$ Myr proposed by \cite{2018A&A...615A..12Y}. The \logg values with this approach are larger than the one derived from spectrum fitting, possibly due to the impact on the {\sc zeeman} estimate of line distortions linked to dark spots, complicated by the heavily blended spectrum (the derivation of \logg relies on the pressure broadened line wings, which are largely hidden by the high \vsini).

Using the K band values of the mass and radius, we derived a theoretical convective turnover time (from the {\sc starevol} models) at one pressure scale height above the base of the convective zone of $25.0 ^{+4.3}_{-3.1}$ days (following the method of \citealt{2016MNRAS.457..580F}). This implies a Rossby number $R_{o}=0.013 \pm 0.002$ with the period obtained by ZDI (see Sec. \ref{sec:DR}). 

A summary of all parameters discussed above can be found in Table \ref{tab:para}. The adopted values for the luminosity, radius, mass, \logg,  and age are taken from our K band calculation, since this band is the least impacted by stellar activity.

\section{Brightness and magnetic field imaging}
\label{sec:ZDI}

\subsection{Multi-line analysis}
\label{subsec:LSD}

The S/N obtained for single spectral lines of V530 Per is too low to extract information about the line profile distortions produced by spots (in \StI) or magnetic fields (in \StV). This situation is especially problematic for V530 Per, due to the rotational broadening resulting in increased blending of lines. We take advantage here of the fact that all photospheric lines mostly display the same shape, with differences from line to line originating from the line depth (in \StI), or from a combination of the line depth, Land\'e factor and wavelength (in \StV). A multi-line approach is therefore a great help to increase the S/N and get rid of the blending issue.   

\begin{figure*} 
   \centering
   \includegraphics[width=19cm,trim={1.0cm 3.0cm 0cm 3.0cm},clip]{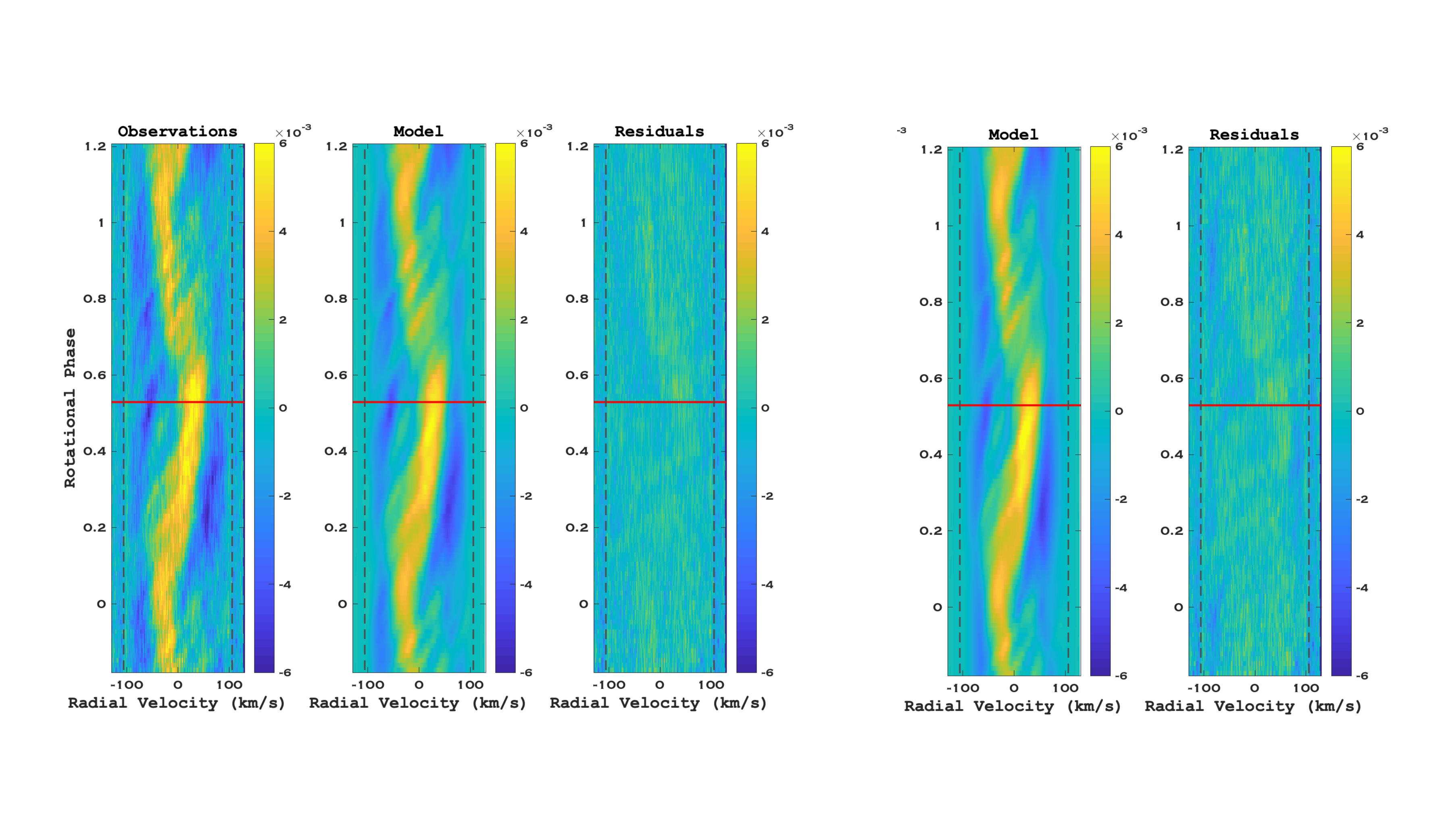}  
\caption{\StI~dynamic spectrum of V530 Per, after subtraction of the averaged \StI~profile. From left to right: observations, DI model with \chit\ $=0.55$ including both bright dark spots, model residuals. The two additional panels on the right show our model with \chit\ $=0.65$ and dark spots only, and the model residual. The vertical, black, dashed lines mark the $\pm$~\vsini\ limit. The portion above the solid, horizontal red lines represents data from 05 Dec 2006, while the lower part represents data taken on 29 Nov 2006.}
   \label{fig:DSb}
\end{figure*}

\begin{figure*} 
\centering
      \includegraphics[width=13cm]{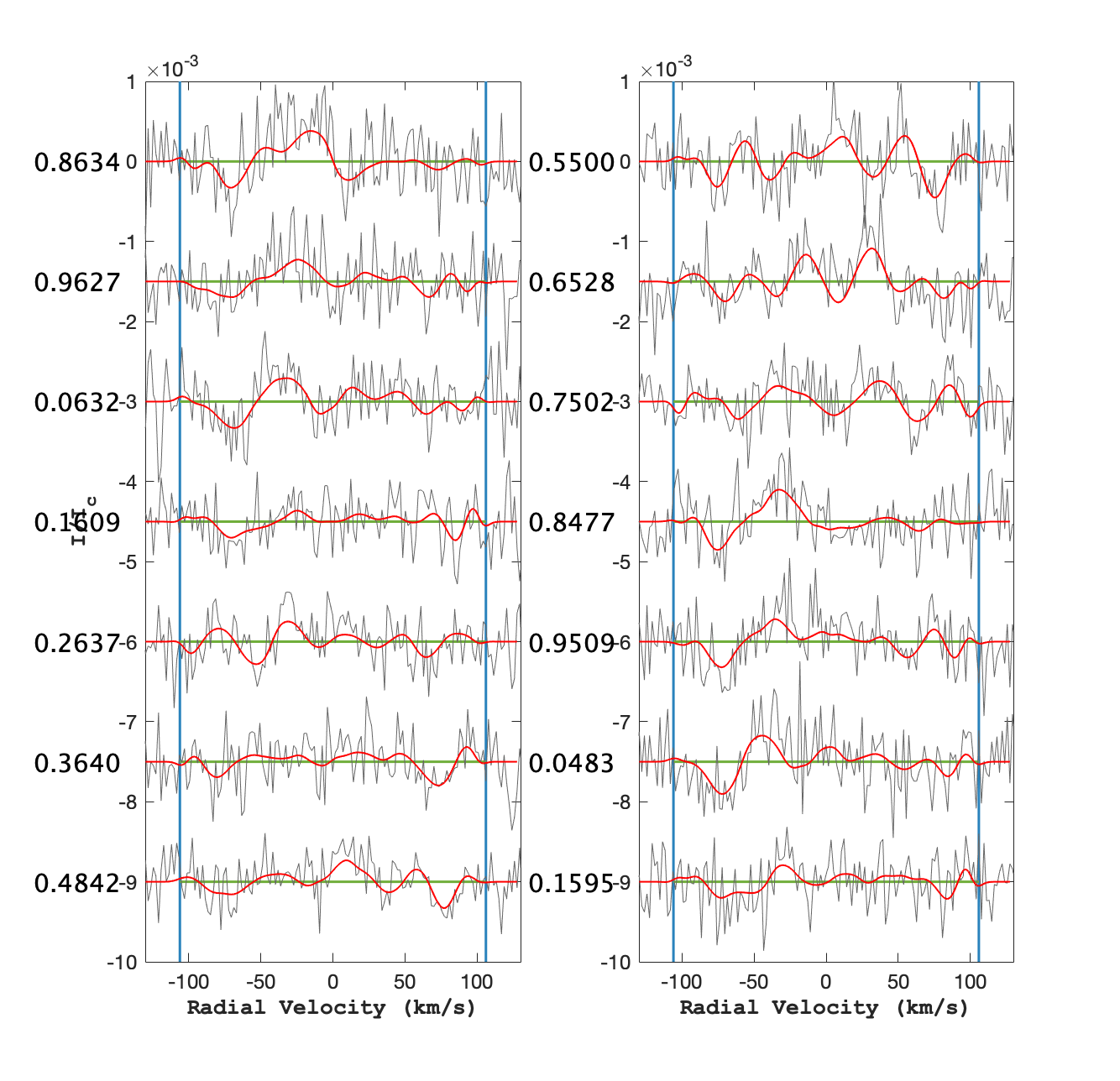} 
      \caption{\StV~LSD profiles of V530 Per. Gray lines represent the observations, while red lines show our ZDI model. Blue vertical lines mark the $\pm$~\vsini\ limit. The left panel shows the data from 29 Nov, and the right panel represents 05 Dec. Rotational phases are indicated on the left of each panel.}
\label{fig:StokesV}
\end{figure*}

All polarized spectra were treated using the Least-Squares Deconvolution method \citep[LSD, see][]{1997MNRAS.291..658D}. This widely employed method computes an average pseudo-line profile from a theoretical list of photospheric spectral lines extracted from the VALD database (getting rid of spectral ranges plagued by telluric or chromospheric lines), using the nearest line list in a grid computed by \cite{2014MNRAS.444.3517M}. Following the fundamental parameters determined in Sec. \ref{sec:param}, we selected a line list with an effective temperature \teff $= 5250 K$ and a logarithmic gravity \logg $= 4.5$, and included in our analysis all lines with a depth greater than 40\% of the continuum level. This resulted in a total of 5726 lines, after removal of all lines plagued by telluric contamination, as well as all lines blended with chromospheric lines. The LSD pseudo-profiles are computed for a velocity step of 1.8 \kms (about $\sim$40\% of the spectral resolution of ESPaDOnS, equal to 4.6~\kms), a normalization Landé factor equal to 1.19, and a normalization wavelength of 650~nm. The resulting set of LSD profiles is plotted in Fig. \ref{fig:DSb} for \StI, and in Fig. \ref{fig:StokesV} for \StV. 

It was noticed by \cite{2016MNRAS.457..580F} that LSD profiles obtained from spectra with S/N below about 70 are sometimes affected by spurious polarized signatures that show up in the polarized line profile and in the Null profile (which is a control parameter that is expected to display only noise). This effect is much less prominent (but sometimes spotted as well) with S/N values between 100 and 150. Given that our observations fall within this second S/N range, we checked that our set of Null LSD profiles were free from any detectable spurious signal, and that the same outcome was reached after averaging all available data together. 

The dynamic spectrum for \StI~shows obvious line distortions which clearly vary with time and self-repeat after one rotation period. This behavior is typical of surface dark spots (producing intensity bumps) or bright regions (translating as intensity dips). The most prominent feature is a systematic bump staying close to the line center, which is indicative of a surface structure anchored at high latitude. The large radial velocity span of this spectral structure tells us that it affects a significant fraction of the visible hemisphere, and its large positive deviation from the average profile reveals a very dark region. Apart from this large spot, a number of smaller trails of both signs in the dynamic spectrum reveal a complex distribution of smaller dark and bright spots. Contrary to the largest spot signature, most of these spectral features can be seen transiting from the blue wing to the red wing of the line profile, but not during their red-to-blue transit (i.e., when they are located behind the visible pole). We can therefore conclude that they are eclipsed during part of the rotation cycle, so that they are likely caused by surface features located at lower latitude than the largest spot which stays visible all the time.    

The time-series of \StV~profiles shows a complex pattern of polarized signatures. We interpret this line polarization as a manifestation of the Zeeman effect. According to the detection criteria proposed by \cite{ 1997MNRAS.291..658D}, only three LSD profiles from 29 Nov. 06 reach the "marginal detection" threshold. All other observations fall in the "no detection" category.The relatively large number of observed rotational phases (and the repeated observation of specific phases) compensates for this poor statistics. Because of the relatively large noise, it is not totally obvious to track progressive changes in the radial velocity and amplitude of the polarized signal. We can however stress that the large polarized feature observed on Nov 29 at phase 0.8634 for negative radial velocities is also observed on Dec 5, at the close-by phase 0.8477. Similar analogies can be observed for other rotational phases that are covered during the two observing nights (e.g., phase 0.0632 versus phase 0.0483, or phase 0.1609 versus phase 0.1595).  

\subsection{Zeeman-Doppler imaging}
\label{sec: ZDIcode}

Our study of V530 Per made use of a new version of the ZDI code described by \citet{2018MNRAS.474.4956F}, which is a Python implementation of the ZDI algorithm presented by \citet{2006MNRAS.370..629D}, based on the maximum entropy fitting routine of \citet{1984MNRAS.211..111S}. The two codes are designed to invert a set of LSD pseudo-line profiles. Using the same data set and identical input parameters, they were shown by \cite{2018MNRAS.474.4956F} to provide nearly identical outcomes. We repeated this test with our own set of observations and reached the same conclusion.

Using \StI~data, the code can compute a DI map of the stellar photosphere assuming that the brightness inhomogeneities are purely generated by dark spots (e.g., \citealt{1997MNRAS.291....1D}), or by a combination of dark and bright patches (e.g., \citealt{2016Natur.534..662D}). The code can also invert \StV~time series to produce a magnetic map, with the additional possibility to use the DI map as a prior assumption. This implementation also includes a basic model of surface differential rotation, that will be detailed in Sec. \ref{sec: df}. 

The surface in the model is divided into a grid of pixels whose edges lie along lines of latitude and longitude with the area of each pixel being roughly the same, as described by, e.g., \cite{1987ApJ...321..496V}. For \StI\ modeling, every pixel is associated with a local line profile, using a simple Gaussian function with a central wavelength taken equal to the normalization wavelength of the LSD profile, a gaussian FWHM taken equal to the one adopted for low \vsini\ stars of the same surface temperature as V530~Per (0.01~nm, \citealt{2016MNRAS.457..580F}), while the gaussian amplitude is equal to 0.533 to match the depth of LSD profiles (a fine-tuning of the line depth is performed with the adjustment of other input parameters, see Sec. \ref{sec: ZDIpara}). Variations in temperature, not  modeled here, can cause small variations in the equivalent width of a line, due to the impact of temperature on local line formation. Generally the impact of small equivalent width variations have a minimal impact on the resulting map (mostly seen as an increase of the best achievable \chit). This rough modeling can be traded for more realistic descriptions of the line shape (e.g., Voigt profiles, \citealt{2018MNRAS.481.5286F}), but given the large \vsini\ of V530 Per its line shape is vastly dominated by rotational broadening, so that a local Gaussian line proves to produce a convincing fitting of the data. For \StV\ modeling, the model considers the brightness map as a prior assumption for the magnetic field reconstruction when \StV~LSD profiles are inverted \cite[e.g., ][]{2014MNRAS.444.3220D}. The magnetic model is computed under the weak field approximation (\StV~is assumed to be proportional to the derivative of \StI), which is a valid approach over the range of field values encountered hereafter (i.e., a few hundred Gauss, e.g., \citealt{2010A&A...524A...5K}). 
The final calculated line profile is a sum of all visible surface elements. In addition to a projection factor depending on the limb angle, and scaling by the brightness map, the continuum is re-scaled to follow a linear limb darkening law of the following form \cite[e.g., ][]{ 2005oasp.book.....G}:

\begin{equation}
    I_{c}/I_{c}^{0} = 1-\eta+\eta \cos(\Phi)
\end{equation}

\noindent where $\eta$ is the limb darkening coefficient, $\Phi$ is the angle from disk center, and $I_{c}/I_{c}^{0}$ is the ratio of local brightness at the limb angle $\Phi$. 
We chose a linear limb darkening coefficient $\eta = 0.73$, by interpolating between available values from the table of \cite{2015A&A...573A..90M}, using  the Kepler filter, as it is the closest in spectral coverage to our instrumental setup, and using stellar parameters of V530 Per given in Sec. \ref{sec:param}. Variations in local line depth relative to the local continuum as a function of limb angle are neglected.

In most available ZDI studies, the stellar surface is assumed to be spherical, implying that the rotational oblateness is neglected. Considering for simplicity the hydrostatic equilibrium in an incompressible, uniform-density star, we obtain to first order (see \citealt{2007ApJ...670L..21T}):

\begin{equation}\frac{R_p}{R_e} \approx 1 - \frac{3\Omega^2}{8G\pi\rho}
\end{equation}

\noindent where $R_p$ and $R_e$ stand for the polar and equatorial radii, $\Omega$ for the rotational rate, $G$ for the gravitational constant and $\rho$ for the density. This rough approach can provide us with an order of magnitude of the oblateness, with $R_p/R_e \approx 0.92$. We therefore modified the local velocities assuming that the stellar surface shape can be described by a Roche model \citep{1978trs..book.....T}, in which the equipotentials $A(R_*,\theta)$ include a centrifugal term:

\begin{equation}
\label{equ:roche}
    A(R_*,\theta)=-\frac{GM_*}{R_*}-\frac{1}{2}\Omega^2 R_*^2 \sin^2\theta
\end{equation}

\noindent where $R_*$ is the stellar radius at  colatitude $\theta$. Whenever $\Omega$ is smaller than the break-up angular velocity $\Omega_{c}$, the surface shape is expressed as follows \citep{1965ApJ...142..265C,1966ApJ...146..152C,1996PhDT........92C}:

\begin{equation}
\label{equ:shape}
    x(\omega,\theta)=\frac{3}{\omega \sin\theta}\cos[\frac{\pi+\cos^{-1}(\omega \sin\theta)}{3}]
\end{equation}

\noindent where $x(\omega,\theta)=R_*(\theta)/R_p$ and $\omega = \Omega / \Omega_{c}$. Taking $\theta=\pi/2$, we can calculate again the oblateness, and obtain $R_p/R_e \approx 0.91$, in agreement with our first approach. Finally, we assume that the surface brightness is affected by gravity darkening. Following the prescription of \cite{1967ZA.....65...89L}, the brightness is therefore assumed to vary as $g^{4b}$, where $g$ is the effective gravity and $b=0.08$, leading to a brightness ratio of 0.88 between the equator and the pole. The exact value that should be used for $b$ is still a matter of debate today (see \citealt{2012A&ARv..20...51V} for a review), but we checked that our model is poorly dependent on the adopted exponent. Changes in the line shape with and without gravity darkening remain modest, as illustrated in Fig. \ref{fig:profiles}. 

To express the magnetic field geometry, the model uses a spherical harmonics decomposition of the field based on \citet{2006MNRAS.370..629D} \citep[see also][for more discussion]{2016MNRAS.459.1533V} by following the set of equations below:

\begin{equation}
B_{r}(\theta,\phi) = \sum_{\ell=1}^{L} \sum_{m=0}^{\ell} \mathrm{Re} [\alpha_{\ell m} Y_{\ell m}(\theta,\phi) ] 
\end{equation}
\begin{equation}
B_{\theta}(\theta,\phi) = -\sum_{\ell=1}^{L} \sum_{m=0}^{\ell} \mathrm{Re}[\beta_{\ell m} Z_{\ell m}(\theta,\phi) +  \gamma_{\ell m} X_{\ell m}(\theta,\phi)] 
\end{equation}
\begin{equation}
B_{\phi}(\theta,\phi) = -\sum_{\ell=1}^{L} \sum_{m=0}^{\ell} \mathrm{Re}[\beta_{\ell m} X_{\ell m}(\theta,\phi) - \gamma_{\ell m}  Z_{\ell m}(\theta,\phi)]
\end{equation}
where:
\begin{equation}
Y_{\ell m} = c_{\ell m} P_{\ell m}(\cos \theta) e^{i m \phi} 
\end{equation}
\begin{equation}
X_{\ell m}(\theta,\phi) = \frac{c_{\ell m}}{\ell+1} \frac{i m}{\sin \theta} P_{\ell m}(\cos \theta) e^{i m \phi} 
\end{equation}
\begin{equation}
Z_{\ell m}(\theta,\phi) = \frac{c_{\ell m}}{\ell+1} \frac{\partial P_{\ell m}(\cos \theta)}{\partial \theta} e^{i m \phi}
\end{equation}
and
\begin{equation}
c_{\ell m} = \sqrt{\frac{2\ell+1}{4\pi} \frac{(\ell-m)!}{(\ell+m)!}}
\end{equation}

Here $(\theta, \phi)$ are the colatitude and longitude on the stellar surface, $P_{lm}$ is the associated Legendre polynomial with $\ell$ and $m$ giving the degree and order of the spherical harmonics mode. In practice, the model is described by a series of three complex coefficients: $\alpha_{\ell m}$ corresponding to the radial poloidal field, $\beta_{\ell m}$ the tangential poloidal field and $\gamma_{\ell m}$ the toroidal field component. We note that the spherical harmonics decomposition is used for both the spherical and oblate stellar geometries described above.

\subsection{ZDI adjustment of stellar parameters}
\label{sec: ZDIpara}

\begin{figure} 
   \centering
   \includegraphics[width=10cm,trim={0.5cm 2.5cm 0 2.5cm},clip]{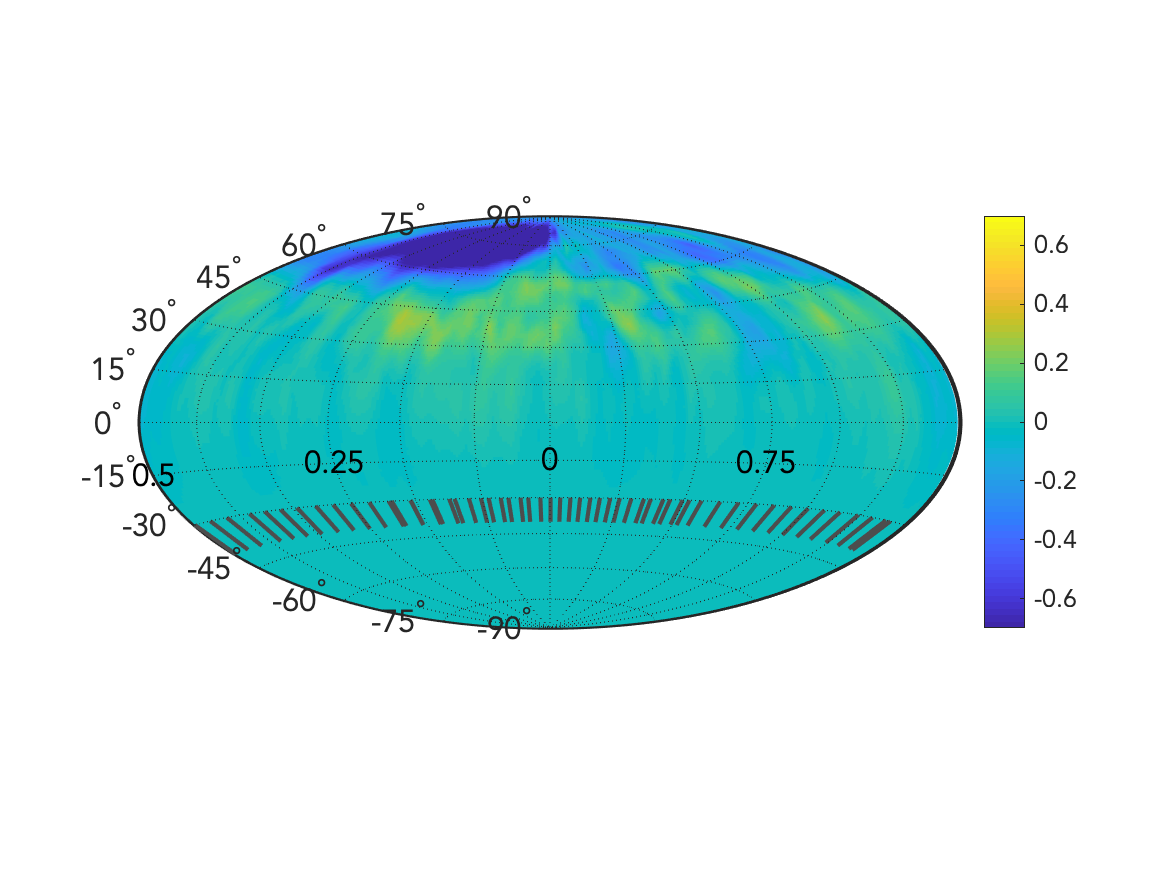}
   \includegraphics[width=10cm,trim={0.5cm 2.5cm 0 2.5cm},clip]{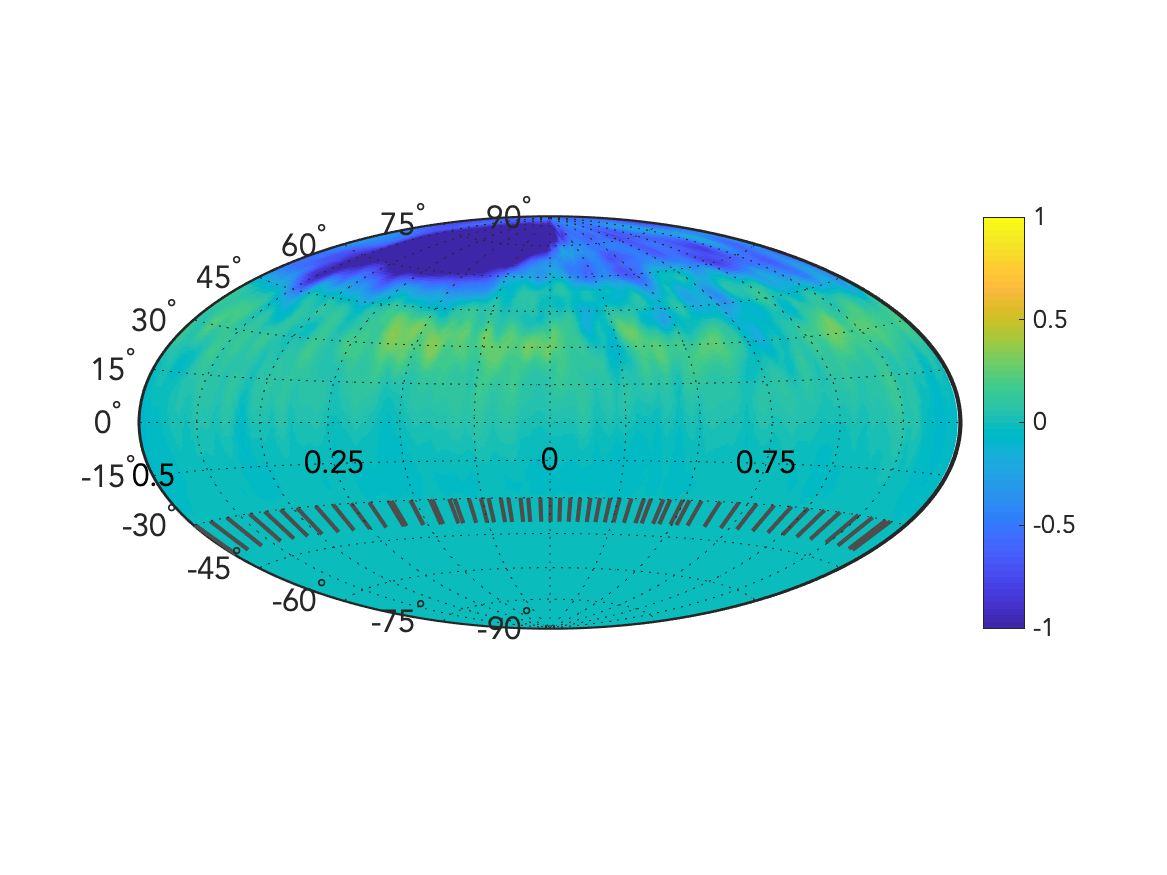}  
   \includegraphics[width=10cm,trim={0.5cm 2.5cm 0 2.5cm},clip]{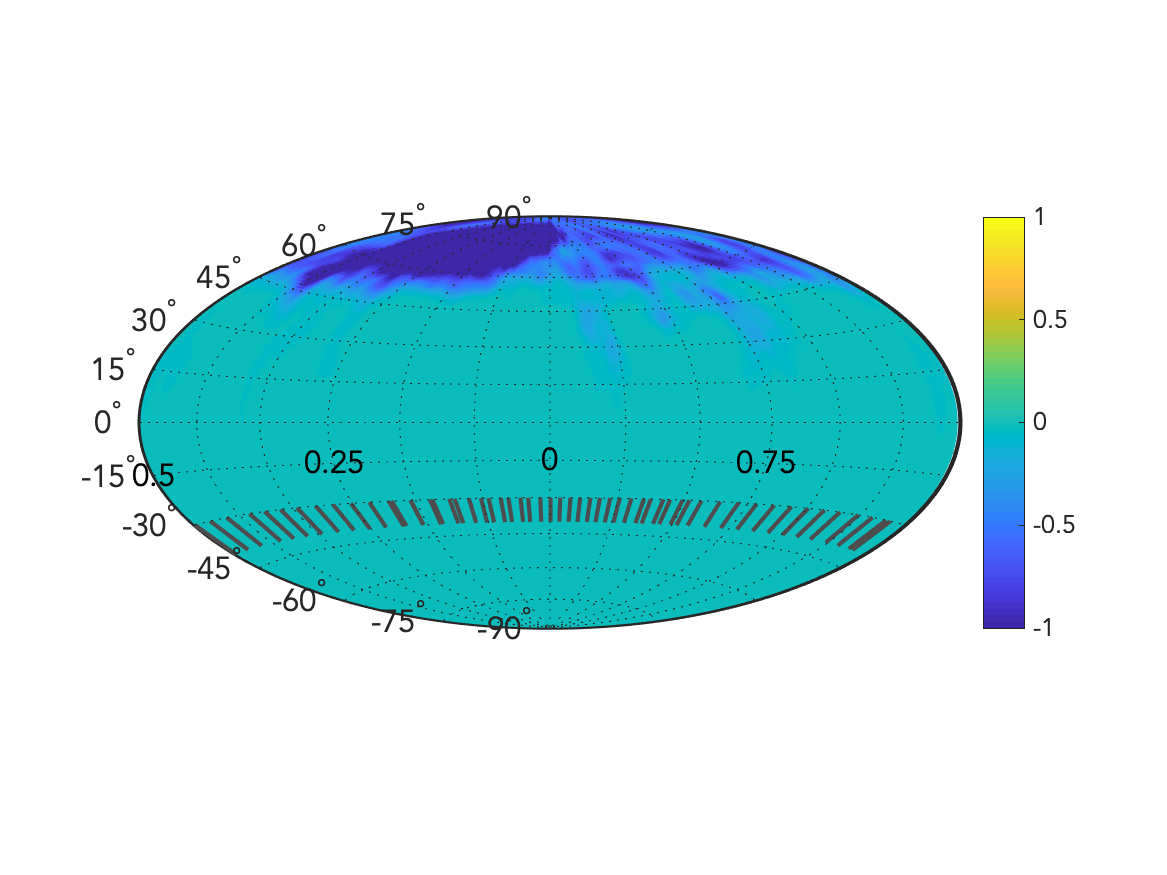}  
\caption{Logarithmic normalized brightness maps of V530 Per reconstructed assuming a combination of dark and brights spots and an spherical surface (upper panel), dark and bright spots and an oblate surface (middle), or dark spots only and an oblate model (lower panel). For display clarity, the gravity darkening was subtracted from the brightess distribution of oblate models. The dark spot model provides us with a reduced  \chit\ $=0.65$, while introducing bright spots reduces this value to 0.55. We used here a Hammer projection of the stellar surface. Meridional ticks in the bottom of the maps mark the rotational phases of our observations. The portion of the maps below $-40\degr$ of latitude is set to 0, as it is invisible to the observer. We emphasize that the color scale is different for spherical and oblate models.}
   \label{fig:brightm_b}
\end{figure}

Stellar parameters extracted from individual spectral lines (Sect. \ref{sec:param}) can be biased when \vsini\ is very large (increasing the number of blends), or when the line profile is distorted by photospheric inhomogeneities. In this case, the optimization of tomographic models can help improve the determination of some parameters, including \vsini, the radial velocity $RV$, the inclination of spin axis $i$, and give access to additional parameters like the rotation period of the equator $P_{\rm eq}$, and the difference in rotational rate between the equator and pole $d\Omega$ (these two parameters will be investigated in Sect. \ref{sec: df}). 

Using a simple \chit\ minimization to determine \vsini\ leads to significant residuals in phase averaged LSD profiles, showing up in the wings of the pseudo-line. Following \cite{2003MNRAS.345.1145D}, we therefore varied again \vsini\ to minimize any systematics (Fig. \ref{fig:residuals}), and finally adopted \vsini\ = 106~\kms (versus 110~\kms\ using \chit\ minimization). This value is slightly larger than the 102~\kms\ obtained by \cite{2001MNRAS.326.1057B}. Although this empirical estimate is not associated to a formal error bar, 1~\kms\ is probably a reasonable order of magnitude for the uncertainty.  

By combining our \vsini\ estimate with the 0.3205 equatorial rotation period of Sect.  \ref{sec: df} and the stellar radius derived in Sect. \ref{sec:param}, we obtained an inclination angle equal to $40 \pm 4^\circ$. Searching for a value of the inclination angle minimizing the \chit\ of the DI model provided us with inconsistent results, depending on whether we used a spherical or oblate model for the stellar shape. While a spherical surface is leading to $i=35^{\circ}$, the oblate model is optimized for values of $i$ below 10$^\circ$, in clear disagreement with other fundamental parameters of V530 Per (this latter value would, for instance, imply an absurdly large stellar radius). The value obtained by \cite{2001MNRAS.326.1057B} was equal to $30^\circ$, in rough agreement with our spherical estimate (assuming a typical error bar, including systematics, on the order of 5 to 10$^\circ$, \citealt{2000A&AS..147..151R}). Facing these discrepant estimates derived from tomographic inversion, we finally adopted $i = 40^\circ$ hereafter.

The \chit\ of the brightness map was minimized for a radial velocity $RV = -0.96\pm 0.04$ \kms, using a total of 7920 freedom degrees (the total number of data points) to estimate the statistical error bar (bearing in mind that instrumental systematics likely dominate this statistical uncertainty, with an absolute $RV$ accuracy probably not better than 0.1~\kms). Although uncertainties were not provided by \cite{2001MNRAS.326.1057B},  our measurement seems to be significantly larger than their estimate ($RV=-3.1$ $\kms$), suggesting that V530 Per is possibly not a single object. We note that the local line depth was fine-tuned after each parameter adjustment (\vsini\ and $RV$), leading to the final value listed in Sec. \ref{sec: ZDIcode}. 

\begin{figure} 
   \includegraphics[width=10cm,trim={1cm 2cm 0cm 1cm},clip]{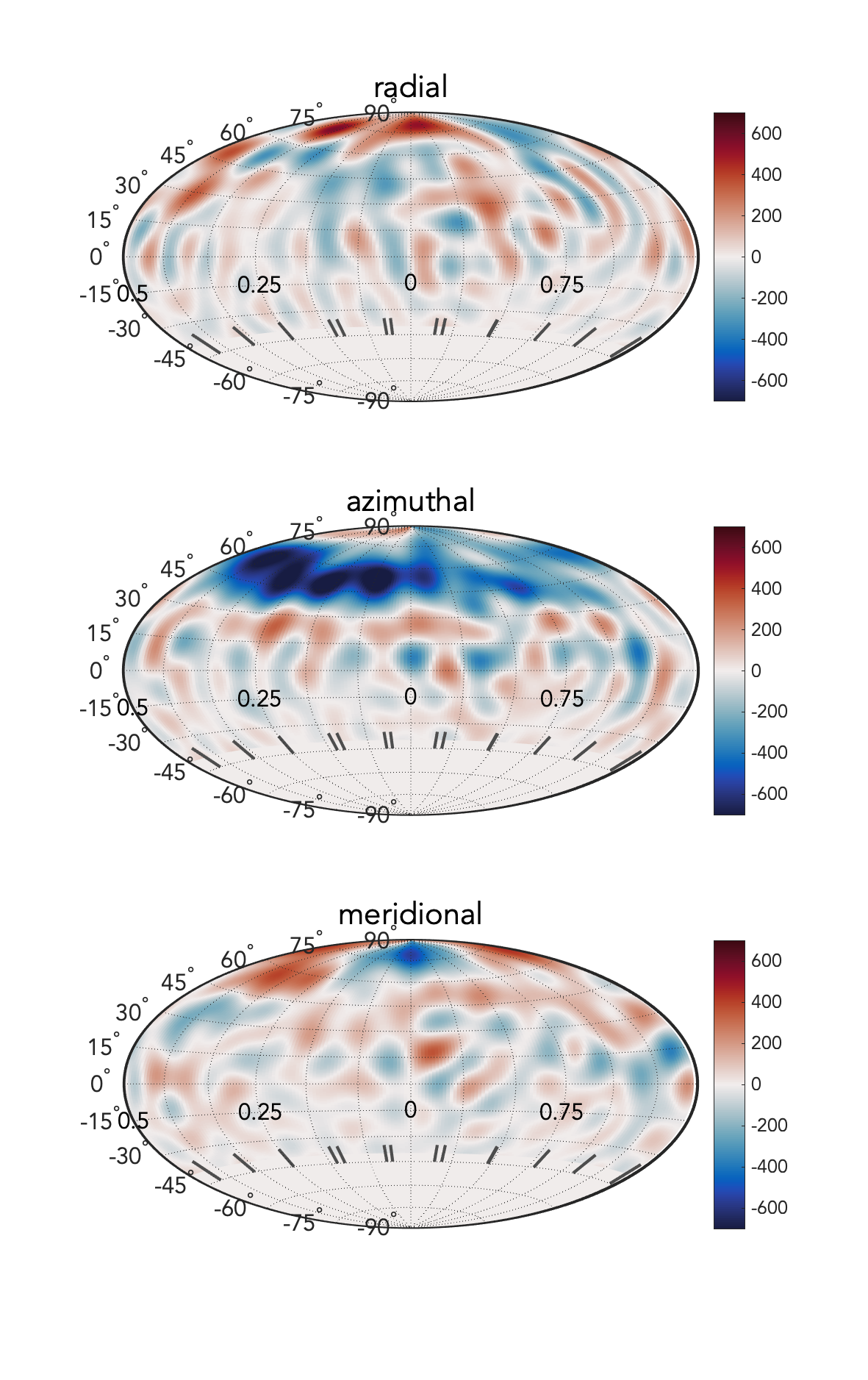}  
\caption{Magnetic map of V530 Per. The three panels show the different field components in spherical projection. The color scale illustrates the field strength in Gauss. A Hammer projection of the stellar surface is adopted, and vertical ticks in the bottom of the panels show the rotational phases of individual \StV~sequences.The portion of the maps below $-40\degr$ of latitude is set to 0, as it is invisible to the observer.}
   \label{fig:magm}
\end{figure}

\subsection{Resulting surface maps}
\label{sec:maps}

\subsubsection{Brightness map}

Thanks to the dense phase coverage described in Section \ref{sec:obs}, we are able to trace the Doppler shifts of surface spot signatures over the stellar rotation period. Stellar parameters determined in Sec. \ref{sec:param}, as well as $d\Omega$ and $P_{\rm eq}$ (Sec. \ref{sec:DR}) are taken as input for the brightness map reconstruction. To the naked eye, the \StI~dynamic spectrum of Fig. \ref{fig:DSb} does not highlight any obvious variations of spot signatures between the two nights of observation, so that we chose to reconstruct the surface brightness distribution using all \StI~data together. 

A first map was reconstructed assuming dark spots only and an oblate surface (lower panel of Fig. \ref{fig:brightm_b}), with a model reaching a reduced \chit\ of 0.65 (showing that error bars in Stokes $I$ LSD pseudo-profiles are over-estimated, as documented by e.g.,  \citealt{2004MNRAS.348.1175P}). The main visible structure is a large spot located around a latitude of $75\degr$, spreading between phases 0 and 0.5. A number of smaller spots are reconstructed as well. Most of the smaller structures are also seen at high latitude, but a few of them are found down to a latitude of $30^{\circ}$. The dynamic spectrum produced with synthetic line profiles of the DI model (see the two panels constituting the right part of Fig. \ref{fig:DSb}) is able to reproduce most observed spectral features. There are, however, some small residuals that reveal the limits of this approach. These remaining signatures are mostly dips, suggesting that they are generated by bright patches (similarly to, e.g., \citealt{2017MNRAS.471..811B}). We therefore computed a second model where both dark and bright spots were allowed, which led us to a smaller reduced \chit\ of 0.55. In spite of the same limb darkening law assumed for dark and bright features, we obtained a nearly flat dynamic spectrum of residuals (third panel of Fig. \ref{fig:DSb}). Although a limb darkening law optimized for dark spots may lead to minor biases in the reconstruction of bright spots, we noticed from the comparison of the two panels in Fig. \ref{fig:brightm_b} that dark spots, at least, do not seem to be noticeably modified by the inclusion of bright regions in the model. The corresponding brightness map shows almost the same distribution of dark spots as in the previous model. Bright spots appear to be concentrated at latitudes lower than the majority of dark spots, with a greater accumulation between 30 and $45\degr$ of latitude. Their size is generally smaller than that of dark spots, and their distribution extends down to equatorial latitudes. The brightest reconstructed spots are about 40\% brighter than the quiet photosphere, which is a higher contrast than observed on the Sun \citep{1979SoPh...63..251H}. The total fraction $Sp_{\rm tot}$ of the stellar surface covered by spots (including both dark and bright features) is equal to 10\% in this model, calculated by the following equation:
\begin{equation}
\centering
   Sp_{\rm tot}=\frac{\sum_{i=0}^{n} |I_{i}-I_{0}|A_{i}}{\sum_{i=0}^{n} A_{i}} 
\end{equation}

\noindent where $I_{0} = 1$ is the brightness with no spot and $I_{i}$ is the brightness on the cell of surface  $A_{i}$.

We note that the brightness map obtained from a spherical model (top panel of Fig. \ref{fig:brightm_b}) is characterized by a lower contrast of the spot pattern (both in the dark, high latitude spots and in the bright, low latitude features). In this case, the fractional spot coverage drops to about 6\%, and the brightest faculae are about 30\% brighter than the quiet photosphere, which remains larger than typical solar values. Another difference compared to the oblate model is a shift of all reconstructed features toward higher latitudes, although this effect is sufficiently subtle to be difficult to distinguish in the maps (the limit of the large polar spot, at latitude $\sim 60\degr$, is where the effect can be most easily seen).

\subsubsection{Magnetic map}
\label{subsec:MagMap}

The magnetic field reconstruction made use of the brightness map as a prior input. The data can be fitted down to \chit\ = 0.9 by including spherical harmonics modes up to $\ell = 15$ (Fig. \ref{fig:StokesV}). Increasing further the number of spherical harmonics coefficients to be fitted does not improve the model. The resulting map displays a complex pattern of magnetic regions, with field strength locally exceeding 1 kG (Fig. \ref{fig:magm}). Other models reconstructed with larger \chit\ values (up to \chit\ = 1) still display  small magnetic features at roughly the same field strength, suggesting that overfitting is not responsible for the observed patchy field distribution. 

The largest and strongest radial field region is reconstructed at the approximate location of the largest dark spot, around a latitude of $75^\circ$. Other radial field spots are more difficult to link to specific brightness patches. The azimuthal field component is dominated by a large belt of negative field encircling the visible pole between latitudes 45 and 60$^\circ$. In this prominent structure, the field reaches a maximum strength between phases 0 and 0.5, which roughly corresponds to the azimuthal location of the largest brightness spot. Its average latitude is comprised between the lower boundary of the polar spot, and the upper boundary of the group of smaller, bright features.   

A list of magnetic parameters was calculated from the $\alpha_{\ell m}$, $\beta_{\ell m}$, and $\gamma_{\ell m}$ spherical harmonics coefficients of the ZDI model to further characterize the magnetic field structure of V530 Per. The resulting list of parameters is shown in Table \ref{tab:magE}. The large ratio between the average magnetic field strength \avB and the unsigned peak magnetic strength $|B_{peak}|$ highlights the complexity of the field structure. We also note that a majority (about two thirds) of the photospheric magnetic energy (as estimated from $B^2$) is stored in the toroidal field component. Focusing on spherical harmonics modes with $m=0$ ($i.e.,$ axisymmetric modes), we note that they contain slightly more than half of the magnetic energy ($\sim53$\%). But a closer look reveals that the poloidal field component is poorly axisymmetric, while the toroidal field energy is mostly reconstructed in axisymmetric structures ($\sim74$\%). As a consequence of the field complexity, a very small fraction of the poloidal magnetic field energy is seen in the dipole ($\ell=1$), quadrupole ($\ell=2$) and octopole ($\ell=3$). Here again, the situation is noticeably different if we consider the toroidal field component where nearly half the magnetic energy ends up in $\ell \le 3$, revealing a higher level of geometrical simplicity in the toroidal field component. 

\begin{figure}
    \centering
    \includegraphics[width=10cm,trim={3.5cm 0cm 0cm 0cm},clip]{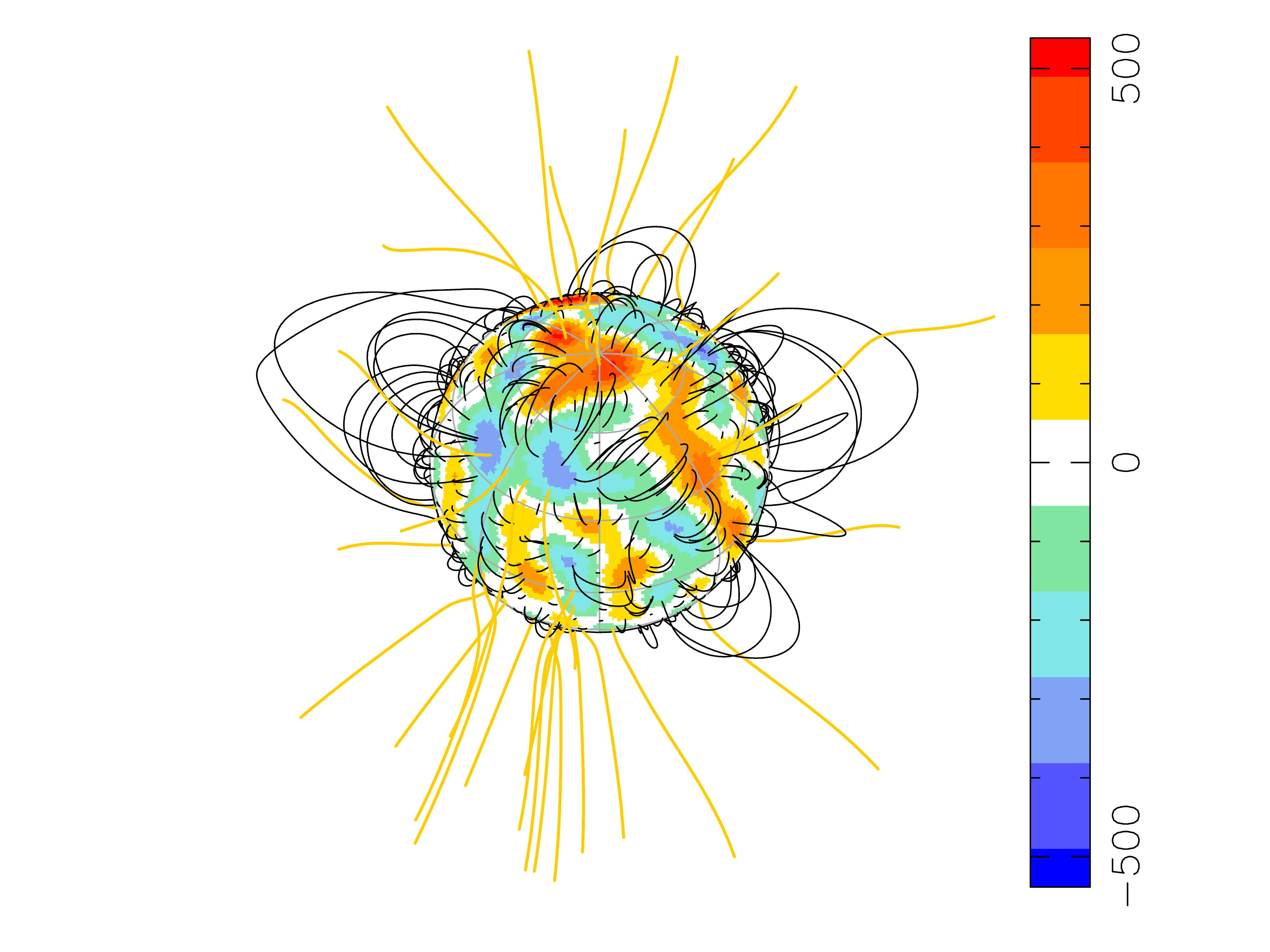}
    \caption{Large-scale potential field extrapolation of V530 Per. The star is seen at phase 0.0, with a $40\degr$ inclination angle. The blue and white lines represent open and close field, respectively. The color scale on the surface show the strength of the radial field component (in Gauss), following Fig.\ref{fig:magm}. The source surface is located at 2.5 $R_{*}$, which is a bit further away from the star than the corotation radius (at 1.9 $R_{*}$).}
    \label{fig:EPPF}
\end{figure}

Finally, we used a potential field source surface model  \citep{2013MNRAS.431..528J} to extrapolate the coronal magnetic field, using the potential component of the ZDI map as boundary conditions (Fig. \ref{fig:EPPF}).  The surface toroidal field component is ignored in the extrapolation, as recent models suggest that purely potential field extrapolations provide a better match to prominence distribution \citep{2020MNRAS.491.4076J}. We assumed that the field becomes purely radial above a source surface located at 2.5 $R_{*}$, which is slightly larger than the corotational radius \citep{2004A&A...414L...5J}. Although the exact location of the source surface is difficult to establish precisely, due to uncertainties in the wind properties, especially in an environment where the centrifugal force plays an important role, we chose this value considering that field lines will likely break open quickly above the corotational radius under the effect of centrifugal forces.

\begin{table}[ht]
    \centering
    \caption{Magnetic field characteristics of AP 149. The values include (a) the average magnetic field strength $\langle B \rangle$, (b) the unsigned peak magnetic field strength $|B_{peak}|$, (c) the ratio of toroidal field energy to the total magnetic energy, (d) the ratio of magnetic energy in axisymmetric modes ($m = 0$) over the total energy, the same quantity but limited to the poloidal (e) and toroidal (f) magnetic component, the ratio of the dipole, quadrupole, and octopole (g, h, i) to the total poloidal component, and ($\ell=1,2,3$) subcomponents of the toroidal field energy, as percentages to the toroidal field energy (j, k, and l).}
    \begin{tabular}{lll}
        \hline
        &Parameter & Value \\
        \hline
(a)&        $\langle B \rangle$  & 177 G \\
(b)&        $|B_{peak}|$  & 1088 G\\
(c)&        toroidal & 64 \%  (total)\\
(d)&        axisymmetric  & 53 \% (total)\\
(e)&        poloidal axisymmetric  & 16 \% (poloidal)\\
(f)&        toroidal axisymmetric  & 74 \% (toroidal)\\
(g)&        dipole ($\ell=1$)  & 1.2 \% (poloidal)\\
(h)&        quadrupole ($\ell=2$) & 3.3 \% (poloidal)\\
(i)&        octopole ($\ell=3$) & 5.4 \% (poloidal)\\
(j)&        toroidal $\ell = 1$  & 8 \% (toroidal)\\
(k)&        toroidal $\ell = 2$  & 21 \% (toroidal)\\
(l)&        toroidal $\ell = 3$ & 20 \% (toroidal)\\
        \hline
        
    \end{tabular}
    \label{tab:magE}
\end{table}


\section{Differential rotation}
\label{sec: df}

\begin{figure} 
   \centering
   \includegraphics[width=9.2cm,trim={0cm 0cm 0cm 0cm},clip]{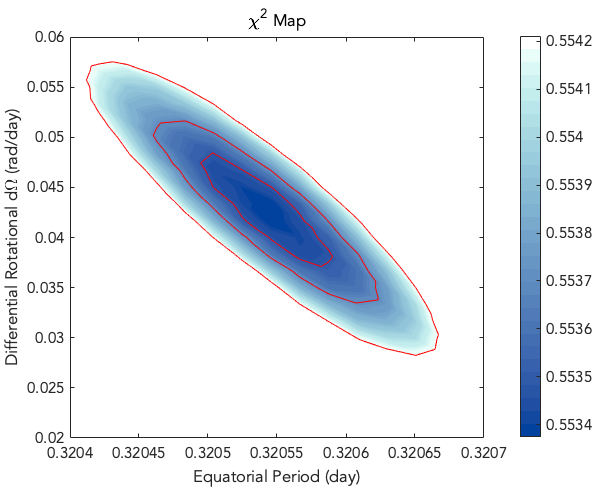}  
\caption{Reduced \chit\ map for the shear parameter $d \Omega$ and the equatorial rotation period $P_{\rm eq}$. The three red solid lines illustrate the $1\sigma$, $2\sigma$, and $3\sigma$ confidence intervals.}
   \label{fig:dOmap}
\end{figure}
\label{sec:DR}

The very dense phase coverage of our time-series, including repeated observations of specific rotational phases separated by a six day gap, constitutes a very good basis to study the short term evolution of photospheric brightness, especially under the action of differential rotation. In this context, the large \vsini\ of V530 Per is also an asset as it gives the capability to spatially resolve small surface features. 

We applied the {\it sheared image} method presented by \cite{2000MNRAS.316..699D} and \cite{2002MNRAS.334..374P}.  This technique incorporates a built-in latitudinal shear of the stellar surface into the DI or ZDI model, and provides robust results even for sparse data sets \citep{2002MNRAS.334..374P, 2003MNRAS.345.1187D}. In our model, the rotation rate $\Omega$ is assumed to vary with the latitude $\theta$, following a simple solar-like dependence:

\begin{equation}
\Omega(\theta) = \Omega_{\mathrm{eq}} - d\Omega \sin^{2}\theta
\end{equation}

\noindent where $\Omega_{\mathrm{eq}}$ is the rotation rate of the equator and $d \Omega$ the pole to equator gradient in rotation rate.

In practice, we estimate the two parameters of this simple law by computing a large number of DI/ZDI models over a grid of values of $d\Omega$ and $\Omega_{eq}$. We then select the doublet that minimizes the model \chit\ (at fixed entropy), as described in Sec. \ref{sec: ZDIpara}. We show in Fig. \ref{fig:dOmap} the \chit\ obtained in the $d\Omega - \Omega_{eq}$ plane using \StI~data, and notice a clear minimum, detected at $P_{eq} = 0.32055 \pm 0.00005$~d and  $d\Omega = 0.042 \pm 0.005$~\rpd. We note that the differential parameters obtained assuming a spherical surface are very close to these values, with $P_{eq} = 0.3205 \pm 0.0001$~d and $d\Omega = 0.045 \pm 0.01$~\rpd. We also inferred the surface shear from a cross-correlation approach similar to the one detailed by \cite{1997MNRAS.291....1D} and used in a number of more recent papers (e.g.,  \citealt{2017MNRAS.471..811B,2019A&A...627A..52K,2019A&A...624A..83K}), and obtain a result (not shown here) in agreement with the sheared image method.

The same procedure was applied to the \StV~data, in order to look for a similar shear of the magnetic geometry. This attempt was not conclusive, most likely because the high relative noise of our polarized line profiles is enough to hide changes in the Zeeman signatures as subtle as those generated by a Sun-like surface shear.

\section{Prominence maps}
\label{sec:prom}

\begin{figure*} 
\centering
      \includegraphics[width=16cm]{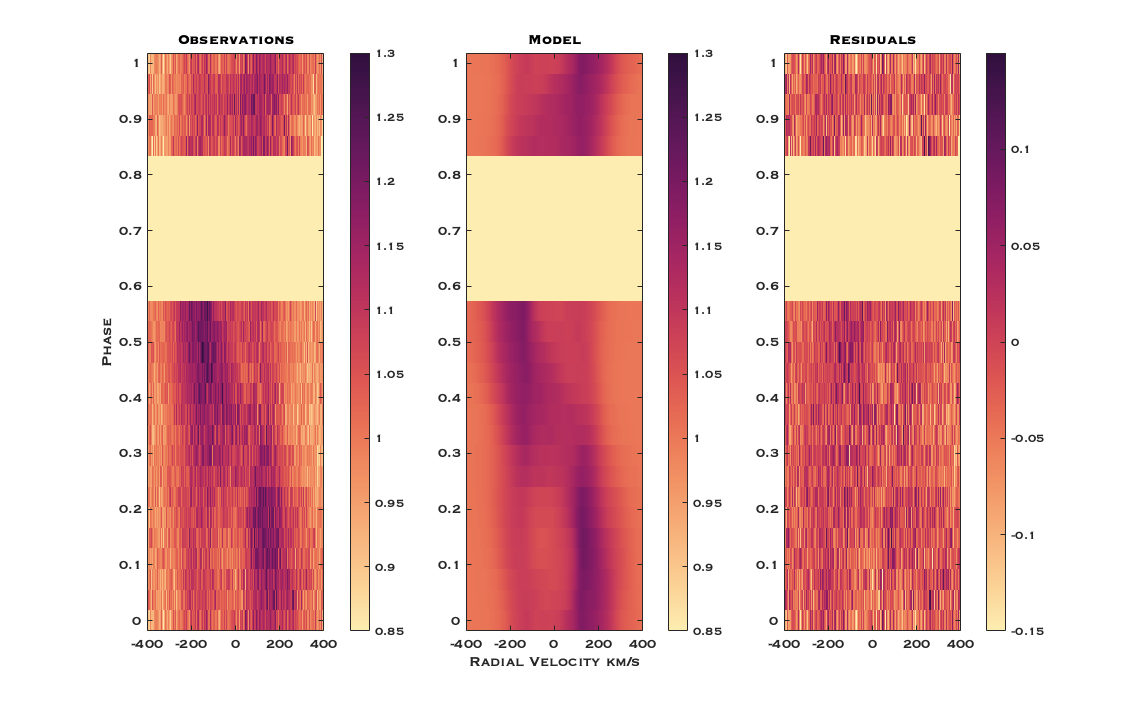} 
      \includegraphics[width=16cm]{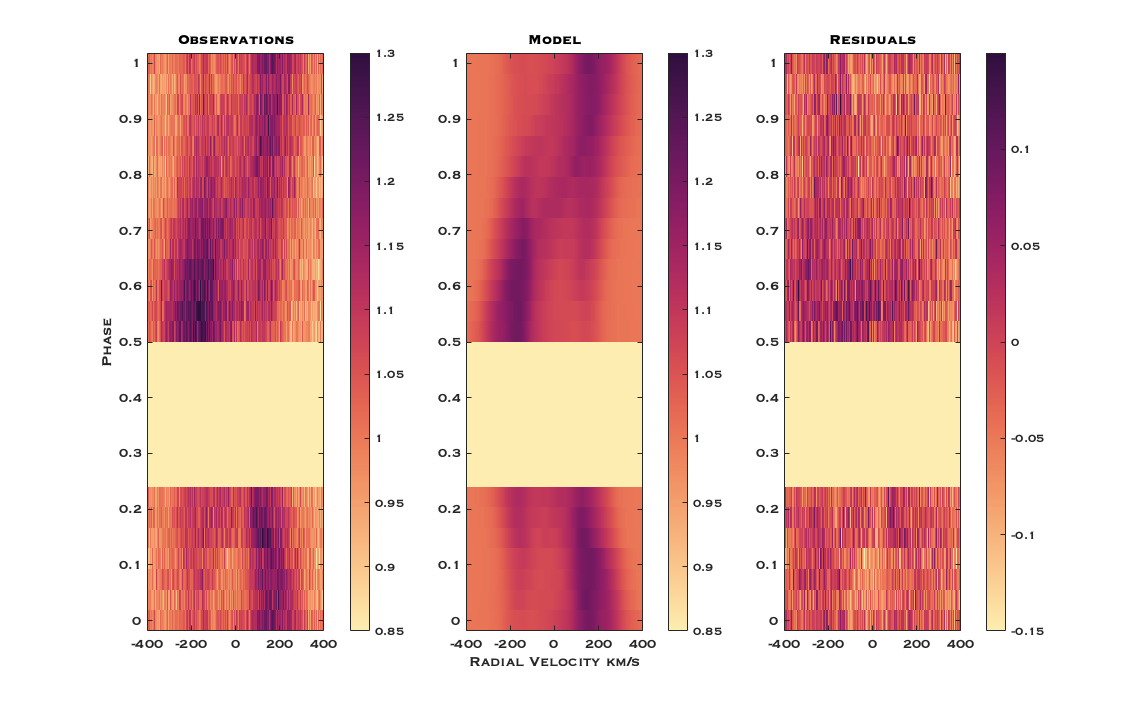}
      \caption{Dynamic Spectra showing the \halpha~line of V530 Per, color coded according to the normalized flux. Rotational phases are computed according to Eq. \ref{equ:emp}. The upper and lower panels are for 29 Nov and 05 Dec, respectively. From left to right, we display the observations, the outcome of the tomographic model, and the residuals. 
      \halpha~mapping from the first night leads to a reduced \chit\ of 7.2, while the second night provides us with a reduced \chit\ of 7.6. }
\label{fig:Prom}
\end{figure*}

\begin{figure} 
   \centering
   \includegraphics[width=9cm]{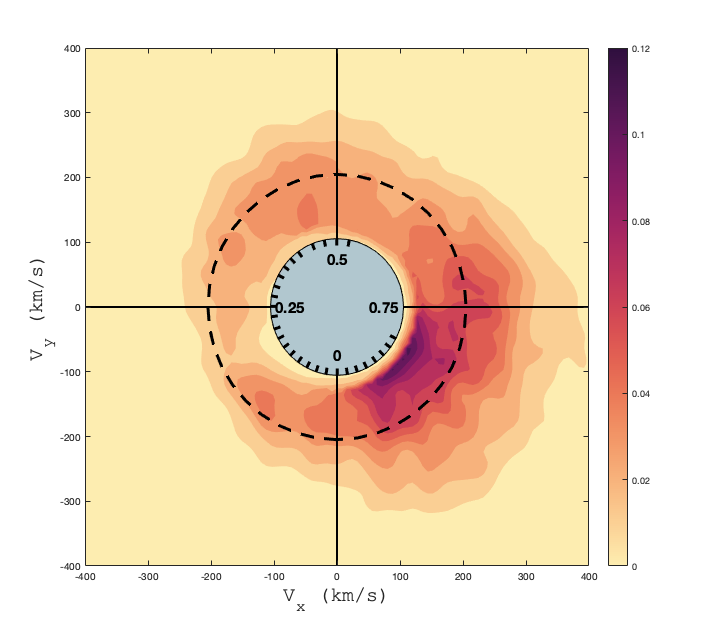}  
   \includegraphics[width=9cm]{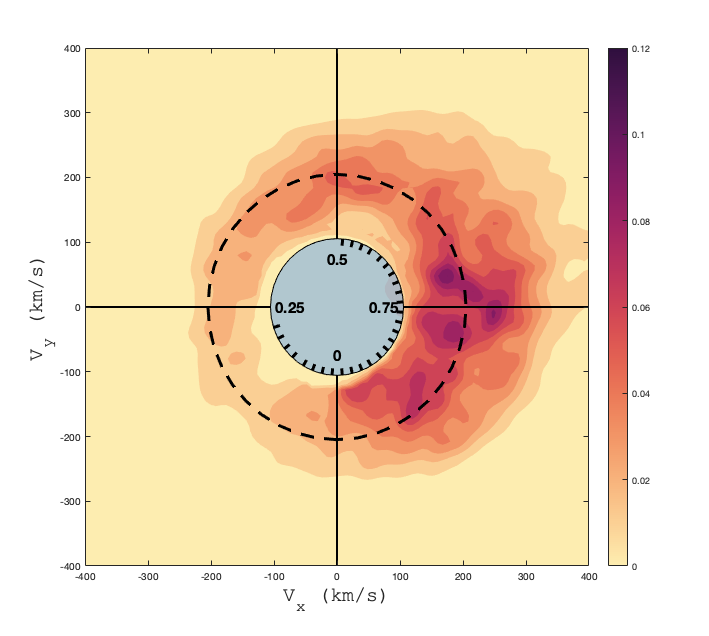}
\caption{Prominence maps of V530 Per reconstructed from the data of 29 Nov, 2006 (top) and 05 Dec, 2006 (bottom). The inner, filled blue circle represents the stellar surface. Radial ticks inside this circle give the rotational phases of our observations. The outer dashed circle is the corotation radius. The color scale depicts the local \halpha~equivalent width, in units of picometers per 8 km~$\rm{s}^{-1}$ square pixel. }
   \label{fig:prom}
\end{figure}

The \halpha~line profiles are always seen in emission throughout the observing run. As shown in the dynamic spectra of \halpha\ in Fig. \ref{fig:Prom}, two peaks are generally observed at roughly $\pm 200$~\kms\ from the line center, and their variations show clear signs of periodic modulation, at a period close to the stellar rotation period. The radial velocities of \halpha\ emitting material is much larger than the projected equatorial velocity of the stellar surface ($\approx 105$~\kms). 

The Alfv\'en radius was calculated with a Weber-Davies model \citep{1967ApJ...148..217W} using the numerical code of \cite{2017A&A...597A..81J}. For this, we have estimated the mass loss from the star using the relation of \cite{2015MNRAS.453.4301S} and the value of the Rossby number in Tab. \ref{tab:para}. We obtain that V530~Per has a mass loss rate of $10^{-10} M_{\odot}.yr^{-1}$. A coronal temperature of $16\times 10^6$ K was calculated applying the relation of \cite{2015A&A...578A.129J}, and $F_x=L_x/(4\pi R_{*})$ with the $L_x$ value quoted in Tab. \ref{tab:para}. We obtain an Alfv\'en radius of about $5R_*$. This value is likely over-estimated, as centrifugal forces are neglected here. It is therefore possible that the actual Alfv\'en radius may be closer to the source surface proposed in Sec. \ref{sec:maps}, although it can be expected to be larger than the source surface radius (e.g.,  \citealt{2003ApJ...590..493S}). Even considering this point, the star is most likely in the centrifugal magnetosphere regime that allows it to support prominences \citep{2008MNRAS.385...97U}. In this case, and as long as sufficiently short time-scales are considered, most of the observed variability can be attributed to the orbital motion of hydrogen clouds. We therefore adopt the assumption that the emission is due to large prominences trapped in the coronal large-scale magnetic field and forced to corotate with the stellar surface \citep{1989MNRAS.236...57C,1989MNRAS.238..657C, 1996MNRAS.281..626S, 2000MNRAS.316..699D}. 

In order to visualize the spatial distribution of prominences, we made use of the simple maximum entropy tomographic code of \citet{2000MNRAS.316..699D}, which is inspired from an algorithm initially developed for cataclysmic variables by \citet{1988MNRAS.235..269M}. Through this inversion method, we attribute 2D Doppler velocities ($V_{x}, V_{y}$) to \halpha~emitting clouds, assuming a local \halpha~profile of gaussian shape with a 0.04~nm ($\sim$18 \kms) gaussian FWHM, under the assumption that the \halpha~emitting material is optically thin \citep{2000MNRAS.316..699D}.  We note that the outcome of the model is mostly unsensitive to the exact value of the local width. In the case of corotating material, the velocity distribution is a straightforward illustration of the spatial distribution. Here, the emission is modeled above the continuum (without subtraction of any reference profile). The reconstructed equivalent widths should therefore not be considered physically meaningful, the main objective of this simple model being to locate the emitting material in the velocity space and highlight its possible short term evolution.

The middle panel of Fig.\ref{fig:Prom} illustrates the synthetic set of \halpha~profiles produced by the code, showing that the main spectral features are correctly reproduced by the model. Residuals, however, remain above the noise level, which shows that the simple model used here is not able to fit the whole \halpha~signal (right panel of the same figure). This mismatch likely highlights that rotational modulation is not the only source of variability in our chromospheric data, even on timescales as short as one night. This observation justifies a posteriori our choice to model the two nights of observation separately. It is also likely that a more elaborated model, e.g., with the intensity of each pixel map allowed to vary sinusoidally with rotation phase (as in, e.g., \citealt{2003MNRAS.344..448S}), would likely allow one to achieve a better fit to the data.  In particular, this additional flexibility may help mimicking, e.g., that some of the prominences may be partly hidden by the star as they rotate.  

The resulting prominence patterns show similarities for the two nights. Large clouds are reconstructed at most phases, at a velocity roughly equal to the one of the corotation radius ($\sim 1.9 R_{*}$, Fig.\ref{fig:prom}). We also note that significant variations are seen between the two nights. In the first map, we reconstructed a higher concentration of material between the stellar surface and the corotational radius, between phases 0.75 and 0.95. During the second night however, the coronal material features a larger radial spread around the corotation radius, as well as azimuthally with a denser material between phases 0.7 and 1. In both cases, most of the \halpha~emission is observed within twice the  corotation radius. We stress, however, that fine details in the observed changes need to be considered with caution, given the very simple model used in our tomographic inversion.

For the two nights, we note that the period optimizing our model is larger than the rotation period used to reconstruct the prominence maps (taken equal to the rotation period of the stellar equator derived in Sec. \ref{sec: df}), with about 0.39~d on 0189 Nov and 0.36~d on 05 Dec. However, such period estimate is likely to be impacted by nonrotational short term variability occurring within each night, as well as by the fact that our data set for each night does not cover a full rotation period.


\section{Discussion}
\label{sec:disc}

\subsection{Brightness map}

The most obvious structure on the brightness map of V530 Per is a very large, dark spot close to the rotation pole (Fig \ref{fig:brightm_b}). This prominent feature is however not axisymmetric, with a latitude of about $75^\circ$. The same recognizable feature was already reported by \cite{2001MNRAS.326.1057B}, using observations collected in 1998. It is therefore plausible that this giant spot is a long-lived structure, or at least an ordinary feature at the surface of V530~Per. 

Large polar spots are commonly observed among the most active stars, most of the time through Doppler mapping as in the present study, but also with interferometry \citep{2016Natur.533..217R}. A number of young solar-type stars at the end of the premain sequence phase or in their early main sequence have been found to host such extended polar caps. Typical examples include {LQ Lup (named RX J1508.6-4423 in the paper}, $P_{\rm{rot}}\sim 0.31d $, Mass$~\sim1.16 M_{\sun}$, \citealt{2000MNRAS.316..699D}) and AB Dor ($P_{\rm{rot}}\sim0.51$ d, Mass$~\sim1  M_{\sun}$, \citealt{1997MNRAS.291....1D,2003MNRAS.345.1187D}). Both of them have a mass and rotation rate similar to those of V530 Per, but both of them display a centered polar spot, while the high-latitude spot observed on V530 Per does not cover the pole. A few other young G dwarfs have been reported to show large off-centered, high-latitude spots (not covering the pole), like EK Dra ($P_{\rm{rot}}\sim2.8$~d, Mass$~\sim0.95 M_{\sun}$,  \citealp{1998A&A...330..685S,2017MNRAS.465.2076W,2018A&A...620A.162J}) or other rapidly rotating members of the $\alpha$~Per cluster (AP 193, He 520, He 699,  \citealt{2001MNRAS.326.1057B}). In V530 Per as in other rapidly rotating stars, the preferential emergence of spots at high latitude can be interpreted as an indication that the Coriolis force can impose magnetic flux tubes to raise toward the stellar surface in a path mostly parallel to the stellar spin axis \citep{1996A&A...314..503S}. Smaller spots are also observed at lower latitudes, which suggests that they may be formed in an internal layer closer to the photosphere.   

A large number of Doppler maps of young stars include dark spots only in their surface model, although some stars have benefited from temperature mapping based on the modeling of individual spectral lines (e.g., V1038~Tau by \citealt{2001A&A...377..264R}, HD~171488 by \citealt{2003A&A...411..595S}, V410 Tau by \citealt{2012A&A...548A..95C}, V1358 Ori by \citealt{2019A&A...627A..52K}, LQ Hya by  \citealt{2019A&A...629A.120C}). Our brightness modeling allows for bright surface patches as well, and the resulting map features a number of bright spots of relatively small area, with a specific concentration between latitudes 30$\degr$ and 45$\degr$. The same set of techniques was also recently applied to the weak line T Tauri stars LkCa 4 \citep{2014MNRAS.444.3220D}, V830 Tau \citep{2016Natur.534..662D}, TAP 26 \citep{2017MNRAS.467.1342Y} and V410 Tau \citep{2019MNRAS.489.5556Y}. These four stars have masses relatively close to the one of V530 Per, although they all are significantly younger, and all rotate slower than V530 Per. We observe a similar mid-latitude accumulation of bright spots on TAP 26 (which has a rotation period about twice that of V530 Per), while other stars from this series do not display this recognizable distribution of bright spots. 

\subsection{Differential Rotation}

The surface shear $d \Omega$ measured for V530 Per is close to the solar value. This estimate is obtained through the assumption that the latitudinal differential rotation follows a smooth solar-like law. This starting point can be questioned in the light of recent numerical simulations, where the most rapidly rotating stars experience a more complex surface rotation pattern in the form of Jupiter-like zonal flows \citep{2017ApJ...836..192B}. The relatively high noise level in our data prevents us from investigating a more complex latitudinal flow pattern in any reliable way. Our simple model tells us at least that solid-body rotation cannot provide us with an optimal model of our time series, and that low latitudes of the visible hemisphere seem to rotate on average faster than the higher latitudes.

A number of rapid rotators benefit from differential rotation measurements obtained with an approach similar to ours. A compilation of these results can be found in \cite{2005MNRAS.357L...1B, 2017MNRAS.471..811B}, highlighting empirical relationships between the surface shear and the rotation rate or surface temperature. Our analysis of V530 Per suggests that the observed shear is roughly in line with previous observations of very rapidly rotating stars of similar $\Omega_{eq}$ values \citep{2005MNRAS.357L...1B}. The observed shear is also in good agreement with the temperature trend reported by the same authors.  

\subsection{Magnetic field}

\begin{figure}
    \centering
    \includegraphics[width = 9.5cm,trim={2cm 0cm 1.5cm 0cm},clip]{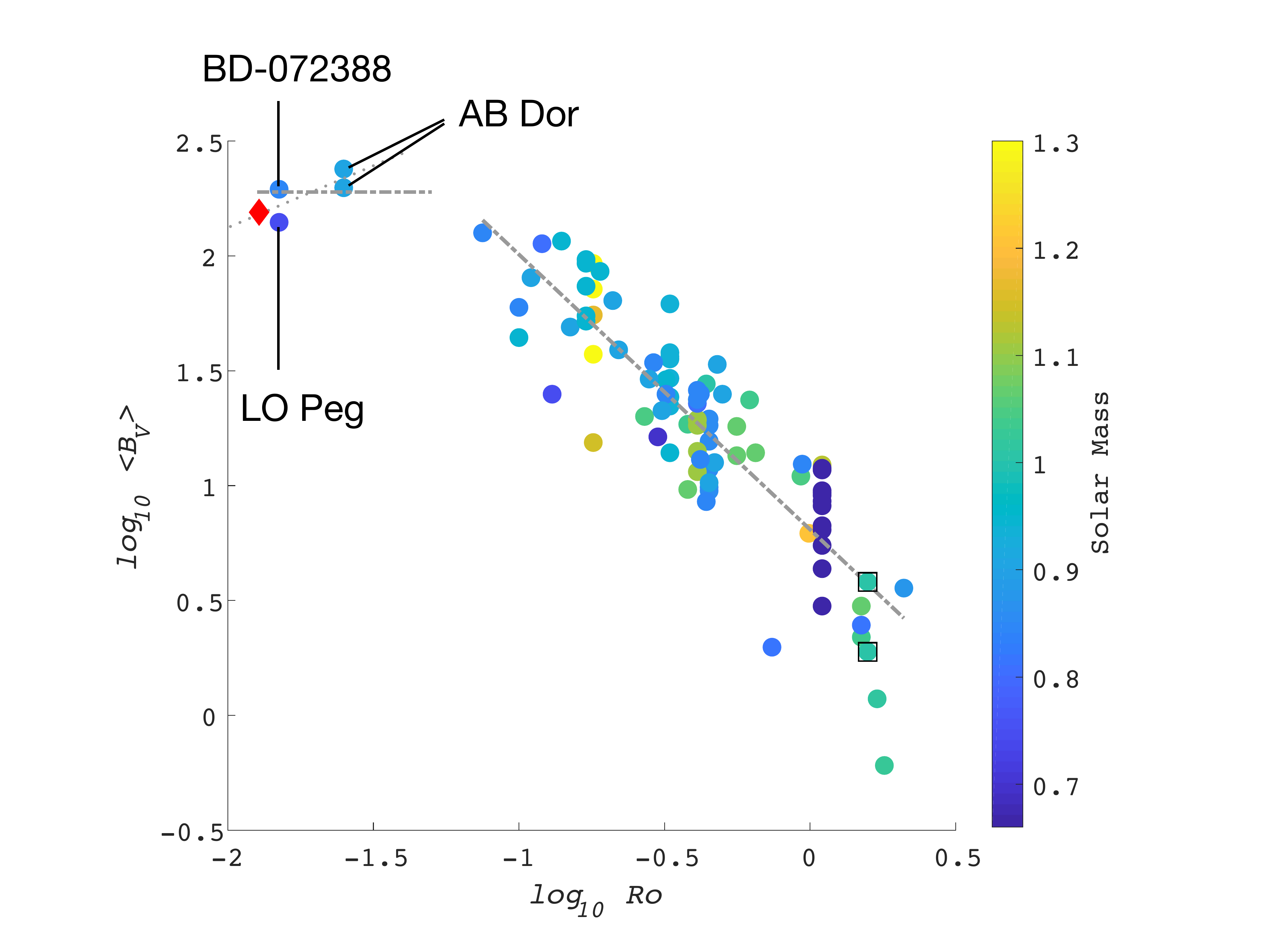}
    \caption{Average large-scale surface magnetic field as a function of the Rossby number for solar-like stars taken from \citet{2019ApJ...876..118S}. Stars are marked by filled circles, color coded according to their mass. Two measurements representative of large-scale field at solar minimum and maximum are marked with squares (and come from \citealt{2014MNRAS.441.2361V}). V530 Per is the red diamond in the upper left corner. Dashed lines show power laws for the saturated and unsaturated regime, with a slope of -1.19 for unsaturated stars and an average \avB~of 186~G for saturated stars. The dotted line is a tentative power law with an exponent of $\approx 0.5$ for saturated stars.}
    \label{fig:RoB}
\end{figure}

The very rapid rotation of V530 Per results in a very small Rossby number, $R_{o} \approx 1.3 \times 10^{-2}$, indicative of a very efficient amplification of its internal magnetic field through a global dynamo. Main sequence stars with similarly small Rossby numbers for which ZDI maps are available generally belong to the M dwarf category, and V530 Per is one of the rare G/K dwarfs populating the low Rossby number branch \citep{2014MNRAS.441.2361V, 2019ApJ...876..118S}. Fig. \ref{fig:RoB} is adapted from the Fig. 1 of \cite{2019ApJ...876..118S} (who present a compilation of ZDI measurements, extending on previous works by \citealt{2014MNRAS.441.2361V,2016MNRAS.457..580F,2018MNRAS.474.4956F}). From the original plot, we have removed M dwarfs because of their much deeper convective envelope. We also discarded Hot Jupiter host stars, since at least $\tau$~Boo was clearly off the main trend, possibly due to significant tidal interaction between the star and its close planetary companion. We therefore  end up with a list of F-G-K effectively single dwarfs. V530 Per stands at the very left of the diagram, making it an important object in the study of saturated dynamo action in ZAMS stars. The average strength of its large-scale magnetic field ($\sim177$ G) is roughly similar to other targets with $\log_{10}(R_{o}) \lesssim -1.5$, with an average field strength of 186 G for this group of five stars. This is in contrast with stars having $\log_{10}(R_{o}) \gtrsim -1.7$, for which the average field strength decreases according to a power law, with \avB~$\propto R_{o}^{-1.19}$. If we plot \avB\ as a function of the rotation period instead of the Rossby number (not shown here), the power law for non saturated stars is such that \avB~$\propto P^{-0.9}$.

We note that the four saturated stars display a marginally decreasing field strength for decreasing Rossby numbers, which could possibly be a hint of supersaturation. This finding would be consistent with X-ray observations of other stars in the $\alpha$~Per cluster, where a decrease of the X-ray flux was reported for stars with the fastest rotation \citep{1996AJ....112.1570P}. A tentative power law fit using the four saturated stars is consistent with \avB~$\propto \sqrt{R_{o}}$. We stress, however, that this trend is based on a small number of objects, and mostly disappears if we include M dwarfs in the same plot or if we assume that the scatter observed a $\log_{10}(R_{o}) \gtrsim -1.7$ (which seems to be mostly due to stellar cycles, e.g., \citealt{2016A&A...594A..29B,2018A&A...620L..11B} for 61 Cygni A) is also valid in the saturated regime.          

The surface field distribution of V530 Per is characterized by a prominent toroidal component, where the majority of the magnetic energy is reconstructed (Table \ref{tab:magE}). This is consistent with the trends reported by \cite{2008MNRAS.388...80P} or \cite{2015MNRAS.453.4301S}, showing that the toroidal field component of cool stars increases faster than the poloidal field when the Rossby number decreases, to the point where the magnetic topology can become dominated by the toroidal component. 

We also observe a very different level of complexity in the toroidal and poloidal field components. The toroidal component has a relatively high fraction of its magnetic energy in low degree modes ($\sim40\%$ in modes with $\ell<4$). It is mostly axisymmetric ($73\%$ in modes with $\ell = 0$), which is consistent with other stars where the toroidal component dominates \citep{2015MNRAS.453.4301S}. The outcome in the magnetic map is a well defined ring of negative azimuthal field. The latitude of this ring ($\sim$50-60$\degr$) is higher than the one of bright spots showing up in the brightness map ( $\sim$35$\degr$) and lower than the latitude of the off-centered polar spot (a polar view grouping the brightness map and the magnetic geometry can be found in Appendix \ref{sec:polar}). We note that the phase of its maximal field strength is close to the phase of the main high-latitude spot ($\sim$0.75). Similar ring-like structures have been identified in other rapid-rotating young dwarfs like AB Dor and LQ Hya \citep{2003MNRAS.345.1145D}, EK Dra \citep{2017MNRAS.465.2076W}, or LO Peg \citep{2016MNRAS.457..580F}. 

In contrast, the geometry of the poloidal field component is much more complex. The dipole, quadrupole and octopole contribute a small fraction of the poloidal magnetic energy ($\le 10\%$ altogether), which is unusual in cool active stars \citep{2016MNRAS.457..580F}. The poloidal field is also highly nonaxisymmetric ($15\%$ in modes with $m=0$). The main radial field region is an extended positive spot covering most of the dark polar spot. The strong radial field reconstructed at high latitude may contribute to generate the dark polar spot, although in other examples of young stars with a giant polar spot, such spatial correlation between the brightness and magnetic geometries is generally not reported (e.g., \citealt{2003MNRAS.345.1145D}).  

\subsection{Prominence system}

The double-peaked emission of the \halpha~line, consistently observed throughout our time-series, agrees well with older observations of \citet{2001MNRAS.326.1057B}. As discussed by these authors, it is likely that the rotationally modulated \halpha\ signal is seen in emission due to the small inclination angle of V530 Per, while stars with a higher inclination feature absorption signatures from their slingshot prominences. Another example of a young star with \halpha\ emitting prominences is LQ Lup \citep{2000MNRAS.316..699D}. \halpha\ transients have been recorded in, e.g.,  AB~Dor \citep{1989MNRAS.236...57C,1997MNRAS.291....1D} and Speedy Mic \citep{2006MNRAS.365..530D,2006MNRAS.373.1308D}.

The tomogram displays multiple clouds corotating with the star and distributed in a ring-like structure at a velocity roughly equal to the one expected at the  corotation radius. This is in striking agreement with similar inversions obtained by \citet{2001MNRAS.326.1057B} for V530 Per, \cite{2000MNRAS.316..699D} for \RXJ and \cite{2006Ap&SS.304...13B} for CC~Eri. Our data suggest that the period of \halpha~emission could be on the order of $0.36 - 0.39$~d, which is longer than the period of the stellar surface, even if we take into account the surface differential rotation. Although the evidence presented here is too slim to reach a definite conclusion about the reality of this longer period (given that our observations in a given night span less than one complete rotation period), it may suggest that the corotating hypothesis is only partially valid, and that prominences sufficiently far away from the surface may experience a less efficient magnetic locking, possibly due to the radial decrease of the field strength. The densest accumulation of prominences shown in the tomogram is located between phases 0.75 and 1. This observation can be linked to the extended, close field lines of Fig. \ref{fig:EPPF} (right-bottom part). We note that the preferred phases for prominences are located away from the phases covered by the large polar spot. Their location is also away from the maximum strength of the radial and azimuthal magnetic field components.   

The dominant part of the \halpha~emission can be modeled by our simple tomographic model, showing that most of the observed system is stable over about 6~hr. Within each night, however, the residuals of the best model highlight nonrotational, short-term changes in the distribution of the coronal material, although we do not witness very fast blue-shifted events similar to those previously reported for AB~Dor  \citep{1989MNRAS.238..657C,1999MNRAS.302..437D}.
The two tomograms display noticeable differences, especially in the azimuthal distribution of the prominence clouds, although the time gap between the two observing nights makes it difficult to say much about the turnover time of the cool coronal material. We note that models of prominence lifetimes in young active stars predict a short life expectancy for stars with a mass and rotation rate similar to V530 Per \citep{2018MNRAS.475L..25V}. A denser temporal monitoring is therefore likely necessary to monitor the short term evolution of slingshot prominences around V530 Per. 

\begin{acknowledgements} 
This work benefited from the support of Programme National de Physique Stellaire (PNPS). TC would like to acknowledge financial support from the China Scholarship Council (CSC). JFD and AAV acknowledges funding from from the European Research Council (ERC) under the H2020 research \& innovation programme (grant agreement \# 740651 NewWorlds and \# 817540 ASTROFLOW). This research made use of the SIMBAD database operated at CDS, Strasbourg, France, and the NASA's Astrophysics Data System Abstract Service. Finally, we thank the referee for insightful suggestions that helped improve this study.
\end{acknowledgements}

\bibliographystyle{aa} 
\bibliography{ap149_draft.bib} 

\appendix

\section{Polar view of the brightness map and magnetic geometry}
\label{sec:polar}

\begin{figure*}
    \centering
    \includegraphics[width=14cm]{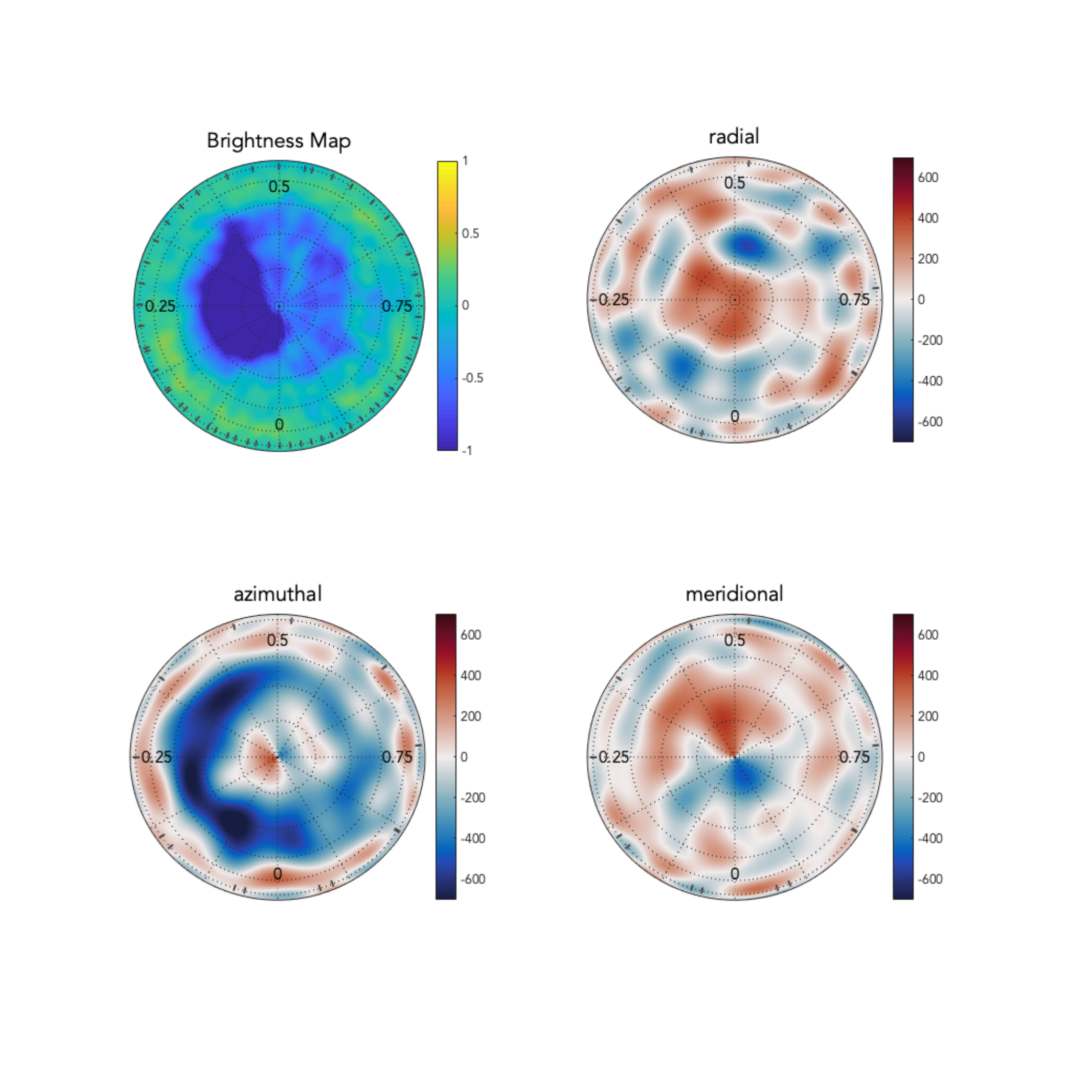}
    \caption{Orthographic polar projection of V530 Per, with the same components and color scale as Fig. \ref{fig:brightm_b} \& \ref{fig:magm}. The latitudinal dash lines represent every 15\degr from 0\degr to 90\degr. Radial ticks close to the equator show the observed phases.}
    \label{fig:ortho_prj}
\end{figure*}

\section{Stokes I LSD profiles}
\label{sec:stokesi}

\begin{figure*}
    \centering
    \includegraphics[width=14cm]{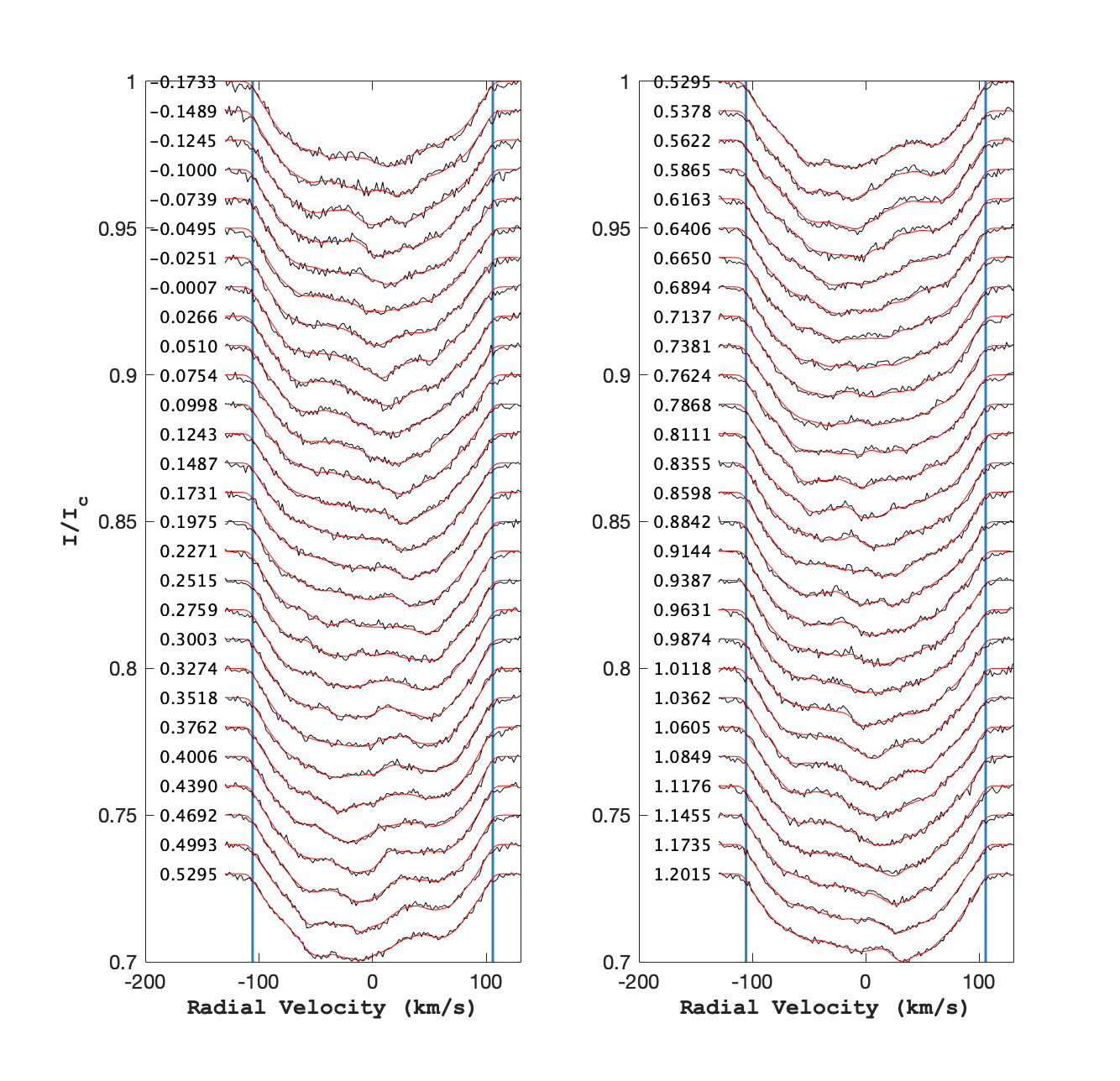}
    \caption{Observed (black) and modeled (red) Stokes I profiles, in a display similar to Fig. \ref{fig:StokesV}. Data from 29 Nov. 06 (resp. 05 Dec 06) is plotted on the left (resp. right). Successive profiles are vertically shifted for display clarity. Rotational phases are indicated on the left of the profiles. Blue vertical lines show the $\pm$ \vsini\ limit.}
    \label{fig:lsd_stI}
\end{figure*}

\section{Spot free line profiles}
\label{sec:profiles}

\begin{figure*}
    \centering
    \includegraphics[width=14cm]{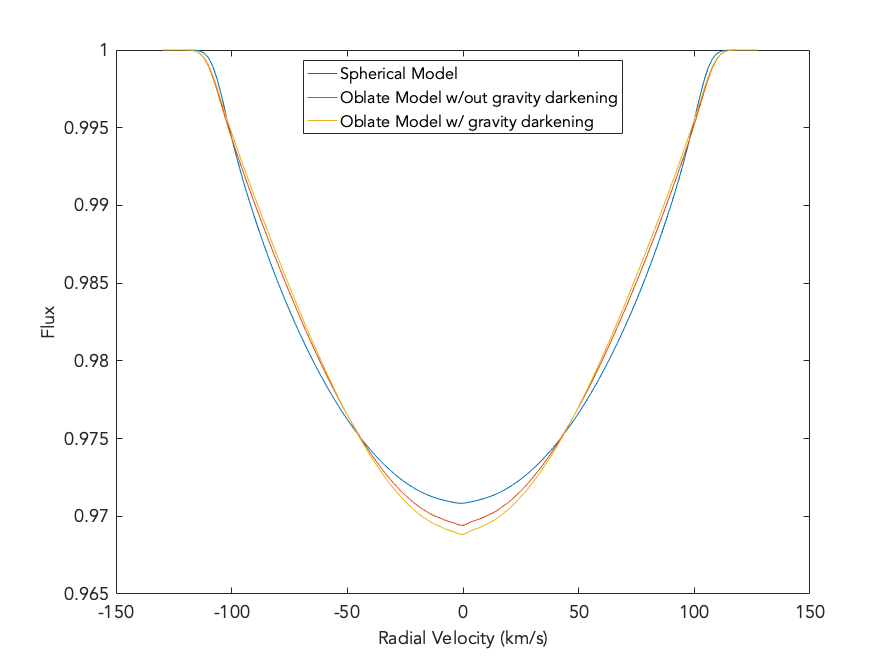}
    \caption{Synthetic, LSD pseudo-line profiles obtained for an unspotted stellar surface assuming a spherical stellar shape (blue), a Roche model (red) and a Roche model with gravity darkening (orange). All other input parameters of ZDI are taken equal in the three models.}
    \label{fig:profiles}
\end{figure*}

\section{DI optimization of the inclination angle}
\label{sec:residuals}

\begin{figure*}
    \centering
    \includegraphics[width=14cm]{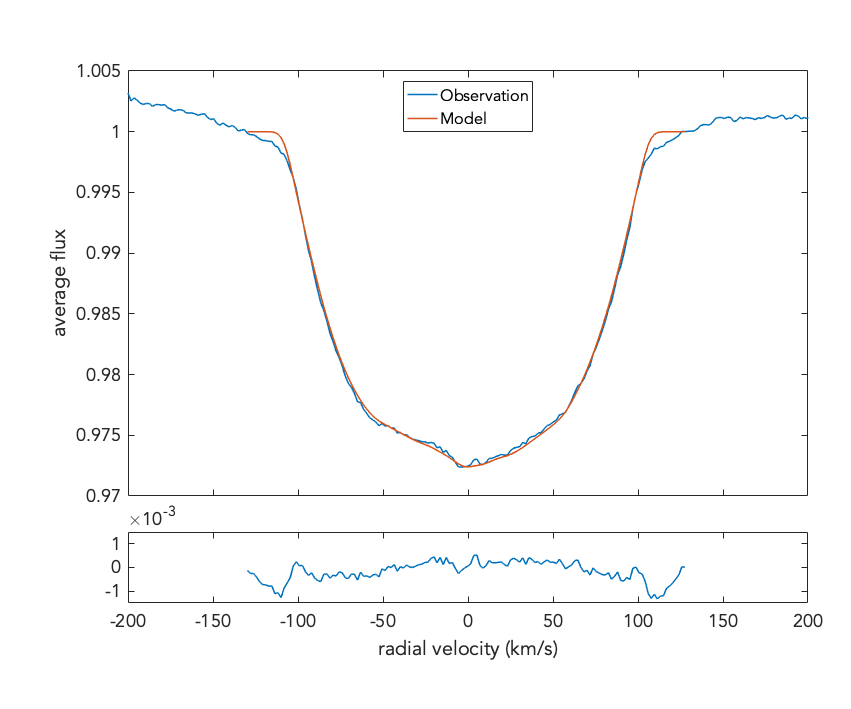}
    \caption{Phase-averaged LSD profile for the V530 Per observations (top panel, blue) and our DI model (top panel, red), as well as the residuals (bottom panel).}
    \label{fig:residuals}
\end{figure*}

\end{document}